\documentclass[twocolumn, twocolappendix]{aastex63}
\usepackage{epstopdf}
\usepackage{graphicx}
\usepackage{natbib}
\usepackage{float}
\usepackage[utf8]{inputenc}
\usepackage{amsmath}
\usepackage{bm}
\usepackage[title,toc,titletoc]{appendix}
\usepackage{outlines}

\newcommand{\project}[1]{{\sffamily #1}}
\newcommand{\emceeplain}{emcee}
\newcommand{\emcee}{\project{\emceeplain}}

\newcommand{\hdust}{\textsc{hdust}}
\newcommand{\angstrom}{\mbox{\normalfont\AA}}

\submitjournal{ApJ}

\begin{document}

\title{The Role of Disk Tearing and Precession in the Observed Variability of Pleione}

\author[0000-0003-0989-2941]{K. C. Marr}
\affil{Department of Physics and Astronomy, University of Western Ontario, London, ON N6A 3K7, Canada}

\author[0000-0001-9900-1000]{C. E. Jones}
\affiliation{Department of Physics and Astronomy, University of Western Ontario, London, ON N6A 3K7, Canada}

\author[0000-0001-7339-4870]{C. Tycner}
\affiliation{Department of Physics, Central Michigan University, Mount Pleasant, MI 48859.}

\author[0000-0002-9369-574X]{A. C. Carciofi}
\affiliation{Instituto de Astronomia, Geofísica e Ciências Atmosféricas, Universidade de São Paulo, Rua do Matão 1226, Cidade Universitária, 05508-900 São Paulo, SP, Brazil}

\author{A. C. Fonseca Silva}
\affiliation{Instituto de Astronomia, Geofísica e Ciências Atmosféricas, Universidade de São Paulo, Rua do Matão 1226, Cidade Universitária, 05508-900 São Paulo, SP, Brazil}

\begin{abstract}

We acquired H$\alpha$ spectroscopic observations from 2005 to 2019 showing Pleione has transitioned from a Be phase to a Be-shell phase during this period. Using the radiative transfer code \hdust\ we created a grid of $\sim100,000$ disk models for Pleione. We successfully reproduced the observed transition with a disk model that varies in inclination while maintaining an equatorial density of $\rho_0(r) = 3\times 10^{-11} (r/R_{eq})^{-2.7}~\rm{g~cm^{-3}}$, and an H$\alpha$ emitting region extending to $15~\rm{R_{eq}}$. We use a precessing disk model to extrapolate the changing disk inclination over $120$ years and follow the variability in archival observations. The best-fit disk model precesses over a line of sight inclination between $\sim25\rm{^{\circ}}$ and $\sim144\rm{^{\circ}}$ with a precessional period of $\sim80.5$ years. Our precessing models match some of the observed variability but fail to reproduce all of the historical data available. Therefore, we propose an ad-hoc model based on our precessing disk model inspired by recent SPH simulations of similar systems, where the disk tears due to the tidal influence of a companion star. In this model, a single disk is slowly tilted to an angle of $30^{\circ}$ from the stellar equator over $34$ years. Then, the disk is torn by the companion's tidal torque, with the outer region separating from the innermost disk. The small inner disk returns to the stellar equator as mass injection remains constant. The outer disk precesses for $\sim15$ years before gradually dissipating. The process repeats every $34$ years and reproduces all trends in Pleione's variability.

\end{abstract}


\section{Introduction} \label{sec:introduction}

Over the past 130 years, the Be star Pleione (HD23862, 28 Tau) has exhibited remarkable spectroscopic and photometric variations. \citet{fro06} first realized the extreme nature of its variability in 1905 and 1906 when he discovered that its previously bright emission lines had disappeared in only a few years. Since this time, Pleione's hydrogen and metal line profiles have exhibited a wide range of shapes, including absorption lines (diskless phase) \citep{mcl38}, doubly-peaked emission lines (Be phase) \citep{doa82} and emission with narrow absorption cores (Be-shell phase) \citep{del73}. The reader can find a historical account of the changes in Pleione’s spectrum by \citet{gul77} and \citet{hir95}, with recent activity described by \citet{nem10}.

The spectral type of B8 Vpe for Pleione was first determined by observations taken during a diskless phase by \citet{lin22}, and was later confirmed by \citet{fro26} and \citet{mer33}. Other details about this system, such as the nature of its companion stars and the interpretation of its variability are much less clear. \citet{kat96a, kat96b} found the near companion to have a $218~\rm{day}$ orbital period with an eccentricity of $0.6$ by analyzing the variation of radial velocities in H$\alpha$ emission from two consecutive shell phases. By similar means, \citet{nem10} confirmed the $218~\rm{day}$ orbit and found that the eccentricity is likely $>0.7$. The inclination angle of the companion's orbital plane has not been reported in the literature.

Long-term variability has also been observed in Pleione since the end of its last diskless phase in $1937$. Since then, its Balmer emission lines have continuously transitioned between Be and Be-shell phases with a $\sim34$ year period \citep{kat96a, hum98}. Using lunar occultation methods, \citet{gie90} detected asymmetry in the disk that was consistent with previously known long-term V/R variability, and speculated that the repeating Be to Be-shell phase transitions are a product of the companion's periastron passages. \citet{ili07} and \citet{ili19} found that the size of the H$\alpha$ and H$\beta$ emitting regions are synchronized with the $\sim34~\rm{year}$ period.

The most recent transition from a Be phase to a Be-shell phase in H$\alpha$ (which, hereafter, will be referred to as the Be phase and the Be-shell phase) from 2006 to 2007 has caused uncertainty about the physical conditions of Pleione's disk causing its variability. Before this transition occurred, \citet{hir07} found the polarization position angle to steadily change over long time-scales. They argued this was the result of a uniformly precessing disk that had been tilted off-axis by tidal interactions with the companion. They asserted that this model explains the repeating Be-shell to Be phase changes, with the shell lines appearing (1973) and disappearing (1988) around a disk inclination of $60^{\circ}$. However, their model was unable to match the rapid drop in magnitude that accompanied the 2006 to 2007 Be to Be-shell transition.

Studying the photometric and spectroscopic variability of Pleione from 2005~November to 2007~April, \citet{tan07a} proposed that the precessing disk suggested by \citet{hir07} had been partially re-accreted before a new disk began building. With this two-disk model, they claim that the newly forming disk explains the formation of new components to the H$\alpha$ and H$\beta$ lines, while the old tilted disk is evidenced by the weakening Balmer line profiles at that time.

Using H$\alpha$ observations spanning from 1994~August to 2009~February, \citet{nem10} argue that the changes in the spectrum cannot be attributed to disk precession, but rather result from physical changes to the disk's structure. Their assertion is that the full width at half maximum (FWHM) of the H$\alpha$ line should increase as the precessing disk becomes more highly inclined. However, they find the FWHM of their H$\alpha$ observations decreases over for a period of $15$ years before the recent Be to Be-shell transition occurred, which they believe indicates physical changes to the circumstellar disk instead of a geometrical effect such as disk precession. Between 2006 and 2007 the FWHM of their H$\alpha$ observations rapidly increased, and they suggest that a new disk has formed.

The goal of this work is to find a Be star disk model that describes the physical and geometrical state of Pleione's disk and evaluate a precessing disk model as constrained by H$\alpha$ spectra, V-band photometry and optical polarimetry. These observations are described along with archival UV through IR continuum observations, as well as photometry and Balmer series spectra corresponding to a diskless phase in Section~\ref{sec:observations}. Section~\ref{sec:stellarparameters} outlines our method for determining the star's parameters. In Section~\ref{sec:results}, we describe our modelling procedures and the best-fit disk models. We also evaluate a precessing disk model and then propose an ad-hoc model that fits all observables based on our findings. Section~\ref{sec:discussion} provides a comparison of our results with the literature and a discussion of future work.

\begin{figure}
	\centering
	\makebox[0.5\textwidth][c]{\includegraphics[width = 0.5\textwidth]{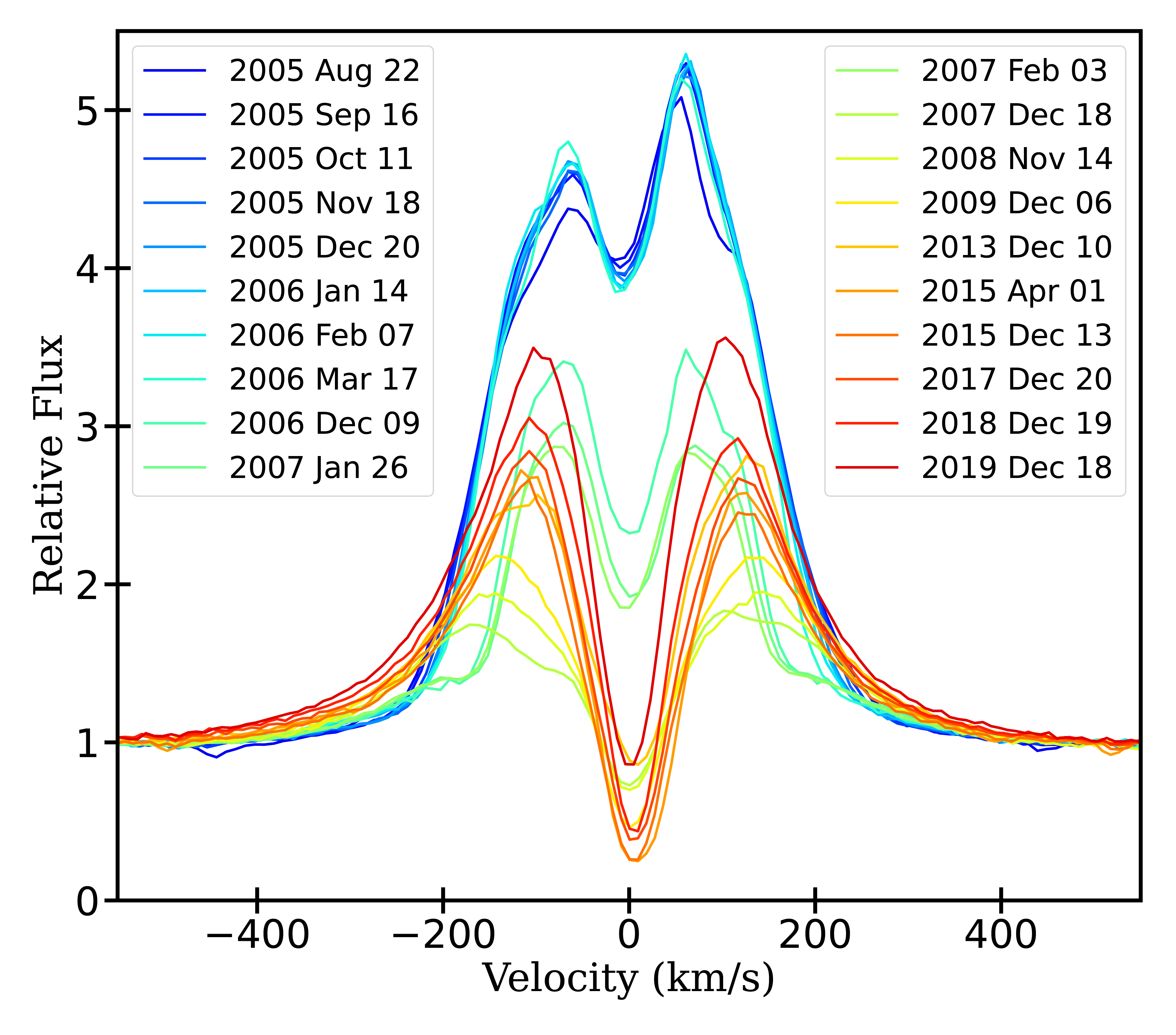}}
	\caption{H$\alpha$ line profile of Pleione observed at Lowell Observatory between 2005~August and 2019~December. The three epochs are differentiated by colour: the Be phase (blue), the transitioning phase (yellow), and the Be-shell phase (red).}
	\label{fig:tycner_halpha_observations}
\end{figure}


\section{Observations} \label{sec:observations}
\subsection{Spectroscopy} \label{subsec:spectroscopy}
\subsubsection{Be and Be-shell Phase Spectra} \label{subsubsec:bebeshellphasespectra}

H$\alpha$ spectra were obtained at the Lowell Observatory, in Flagstaff, AZ, USA, using a fiber-fed echelle spectrograph connected to the 1.1~m John S. Hall telescope. The properties of the spectrograph are outlined in \citet{hal95} and the spectroscopic reduction steps used to extract the spectra from echelle orders are described in \citet{hal94}. We have acquired 31 spectra with this instrument from 2005 to 2019 at a resolving power of 10,000 in the H$\alpha$ region. These observations are available online in a machine readable format (Table~\ref{tab:tycner_halpha_obs}). 

The H$\alpha$ spectra shown in Figure \ref{fig:tycner_halpha_observations} begin during Pleione's most recent Be phase in 2005 and follow the transition to a Be-shell phase in 2006 and 2007. From 2008 to 2015, the line brightened as the ratio of the peak flux over the continuum flux~($F/F_c$) increased from $\sim2$ to $2.5$. During 2015, the system's H$\alpha$ equivalent width (EW) dropped by $\sim20~\rm{\%}$. From 2016 until our latest observations in 2019, the line profile has continued to brighten to a maximum of $F/F_c \approx 3.5$.

Table \ref{tab:tycner_halpha_characteristics} summarizes the characteristics of each spectrum. In particular, the peak flux of the H$\alpha$ profile was on average $F/F_c \approx 5.2$ from 2005~August to 2006~March. After the transition from the Be to Be shell phase between 2006~March and 2006~December, the peak flux had decreased to a minimum of $\sim1.8$. By 2013, the peak flux increased to $\sim2.8$, and, despite a small decrease in 2015, continued to increase to $\sim3.5$ by 2019~December.

The violet (V) and red (R) peaks, measured as the maximum flux values within each peak, had the greatest difference during the Be phase, between 2005~August and 2006~February, with an average ratio of $V/R \approx 0.87$. Afterward, during the shell phase, the $V/R$ ratio remained closer to unity.

The EW was relatively constant between 2005~August and 2006~March. Between 2006~March and 2006~December, during the transition from the Be to Be-shell phase, the H$\alpha$ EW decreased by $\sim1.1~\rm{nm}$. The EW increased by $\sim0.5~\rm{nm}$ between 2008 and 2013, and then decreased by $\sim0.3~\rm{nm}$ by the end of 2015. Since then, the EW has increased by $\sim0.9~\rm{nm}$ to $-1.7~\rm{nm}$.

Archival H$\alpha$ spectra of Pleione from the Be Star Spectra Database\footnote{http://basebe.obspm.fr/basebe/} were also acquired to track the change in H$\alpha$ EW over time. We used 419 spectra from 2007 to 2021 observed by amateur astronomers Ernst Pollmann \citep{pol20} and Joan Guarro i Fl\'o. We also used H$\alpha$ EW observations from \citet{hir76, hir95, hir07}, which collectively span from 1953 to 2004. Each of these measurements are presented alongside our own for comparison to our precessing disk model in Section \ref{sec:results}.

\begin{figure}
	\centering
	\makebox[0.5\textwidth][c]{\includegraphics[width = 0.5\textwidth]{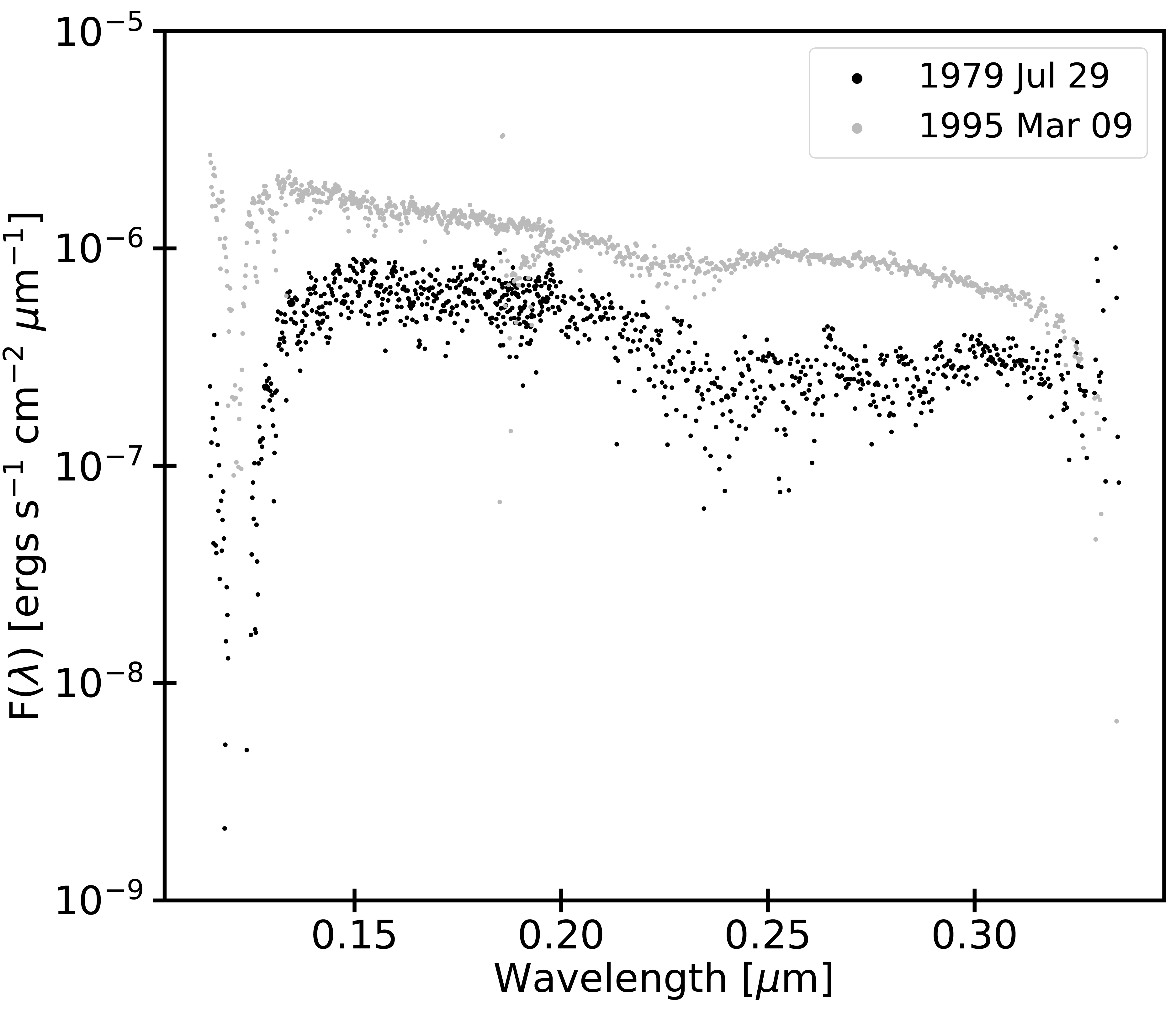}}
	\caption{Upper and lower limits of the variable UV continuum observed by the IUE.}
	\label{fig:IUE_spectra}
\end{figure}

We obtained 70 ultraviolet (UV) spectra, observed by the International Ultraviolet Explorer (IUE) between 1979~July and 1995~March, from the INES database \citep{wam01} following the selection procedure described by \citet{fer12}. These observations were taken with the large aperture and high dispersion settings on both the short-wavelength (1,150 -- 2,000 $\rm{\angstrom}$) and long-wavelength (1,850 -- 3,300~\rm{\angstrom}) spectrographs to ensure proper flux calibration and a high spectral resolution of approximately $0.2~\rm{\angstrom}$ \citep{esa00}. We chose to remove the IUE data beyond $0.3~\rm{\micron}$ due to instrumental limitations that cause significant uncertainty \citep{wam01}. Figure \ref{fig:IUE_spectra} illustrates how the UV flux steadily increased by a factor of $\sim3$ to $4$ during the time-frame of the IUE observations. How this affects our determination of the stellar parameters is discussed in Section \ref{sec:stellarparameters}

\begin{table}
	\centering
		\caption{H$\alpha$ observations of Pleione from the Lowell Observatory. The 31 spectra are identified chronologically by spectrum ID number. The full table is available online in machine-readable form.}
	\begin{tabular}{cccc}
		\hline
		\hline
		Spectrum & MJD & $\lambda$ & $F/F_c$  \\
		ID Number &  (+2400000.5) & [nm] & \\
		\hline
	
        1	&	53604	&	648.183655	&	1.005251	\\
        1	&	53604	&	648.207947	&	1.036522	\\
        1	&	53604	&	648.232300	&	0.971167	\\
           
		\hline
		\hline
	\end{tabular}
	\label{tab:tycner_halpha_obs}
\end{table}

\begin{table}
    \centering
	\caption{H$\alpha$ emission line profile characteristics of Pleione.}
	\begin{tabular}{ccccc}
		\hline
		\hline
		Date & MJD & EW & Peak & V/R  \\
		 & (+2400000.5) & [nm] & Flux & Ratio  \\
		\hline
        2005 Aug 22	& 53604 & -2.571	&	5.081	&	0.861	\\
        2005 Sep 16	& 53629 & -2.691	&	5.296	&	0.867	\\
        2005 Sep 17	& 53630 & -2.691	&	5.353	&	0.850	\\
        2005 Oct 11	& 53654 & -2.658	&	5.291	&	0.871	\\
        2005 Nov 18	& 53692 & -2.619	&	5.298	&	0.871	\\
        2005 Dec 20	& 53724 & -2.620	&	5.215	&	0.896	\\
        2006 Jan 14	& 53749 & -2.630	&	5.308	&	0.878	\\
        2006 Jan 24	& 53759 & -2.562	&	5.122	&	0.874	\\
        2006 Feb 07	& 53773 & -2.650	&	5.355	&	0.871	\\
        2006 Mar 17	& 53811 & -2.563	&	5.198	&	0.924	\\
        2006 Dec 09	& 54078 & -1.458	&	3.484	&	0.979	\\
        2007 Jan 26	& 54126 & -1.209	&	3.021	&	1.050	\\
        2007 Feb 03	& 54134 & -1.160	&	2.870	&	1.012	\\
        2007 Dec 18	& 54452 & -0.621	&	1.833	&	0.952	\\
        2008 Nov 14	& 54784 & -0.676	&	1.955	&	0.994	\\
        2008 Nov 15	& 54785 & -0.704	&	1.985	&	0.985	\\
        2008 Nov 15	& 54785 & -0.663	&	1.952	&	0.996	\\
        2009 Dec 06	& 54994 & -0.825	&	2.180	&	1.005	\\
        2013 Dec 10	& 56636 & -1.174	&	2.804	&	0.916	\\
        2013 Dec 10	& 56636 & -1.194	&	2.811	&	0.913	\\
        2013 Dec 11	& 56637 & -1.160	&	2.824	&	0.900	\\
        2015 Apr 01	& 57113 & -0.941	&	2.725	&	1.052	\\
        2015 Apr 01	& 57113 & -0.928	&	2.721	&	1.055	\\
        2015 Dec 13	& 57369 & -0.860	&	2.656	&	1.083	\\
        2015 Dec 14	& 57370 & -0.862	&	2.661	&	1.084	\\
        2017 Dec 20	& 58107 & -1.097	&	2.842	&	1.064	\\
        2017 Dec 20	& 58107 & -1.100	&	2.854	&	1.067	\\
        2017 Dec 20	& 58107 & -1.097	&	2.854	&	1.072	\\
        2017 Dec 21	& 58108 & -1.082	&	2.843	&	1.070	\\
        2018 Dec 19	& 58471 & -1.325	&	3.053	&	1.044	\\
        2018 Dec 20	& 58472 & -1.381	&	3.132	&	1.058	\\
        2019 Dec 18	& 58835 & -1.748	&	3.560	&	0.982	\\
        2019 Dec 18	& 58835 & -1.795	&	3.582	&	0.982	\\
        2019 Dec 19	& 58836 & -1.723	&	3.517	&	0.986	\\
        2019 Dec 19	& 58836 & -1.718	&	3.506	&	0.991	\\
		\hline
		\hline
	\end{tabular}
	\label{tab:tycner_halpha_characteristics}
\end{table}

\subsubsection{Digitization of Diskless Spectra}\label{subsubsec:digitizationofdisklessspectra}

\begin{figure}
	\centering
	\makebox[0.5\textwidth][c]{\includegraphics[width = 0.4\textwidth]{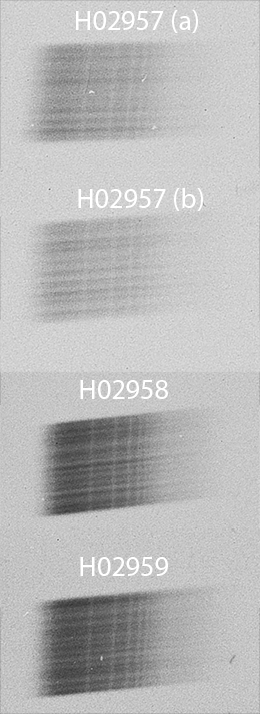}}
	\caption{Photographs of Balmer series observations of Pleione taken on photographic plates in 1927 at the Cambridge Observatory. Each spectrum contains H$\alpha$, H$\beta$, H$\gamma$, and H$\delta$ lines in order from left to right. These line profiles were obtained during Pleione's last diskless phase. The plates are labelled for each spectrum by their designation in the Harvard Astronomical Plate Collection.}
	\label{fig:photographs_of_plates}
\end{figure}

\begin{table}
    \centering
    \caption{Stellar log of observations that contained diskless spectra of Pleione.}
	\begin{tabular}{ccccc}
		\hline
		\hline
		Date & Plate & Exposure & Prism & Exp. Time \\
		 & Designation & Number &   & [min] \\
		\hline
        1927 Sep 26	&	0 & H02957	& D	&	3	\\
        1927 Sep 26	&	1 & H02957	& D	&	3	\\
        1927 Sep 27	&	1 & H02958	& 6$^{\circ}$	&	4	\\
        1927 Sep 27	&	1 & H02959	& 6$^{\circ}$	&	4	\\
		\hline
		\hline
	\end{tabular}
    \label{tab:harvard_plates}
\end{table}

\begin{figure}
\centering
\includegraphics[width=\columnwidth]{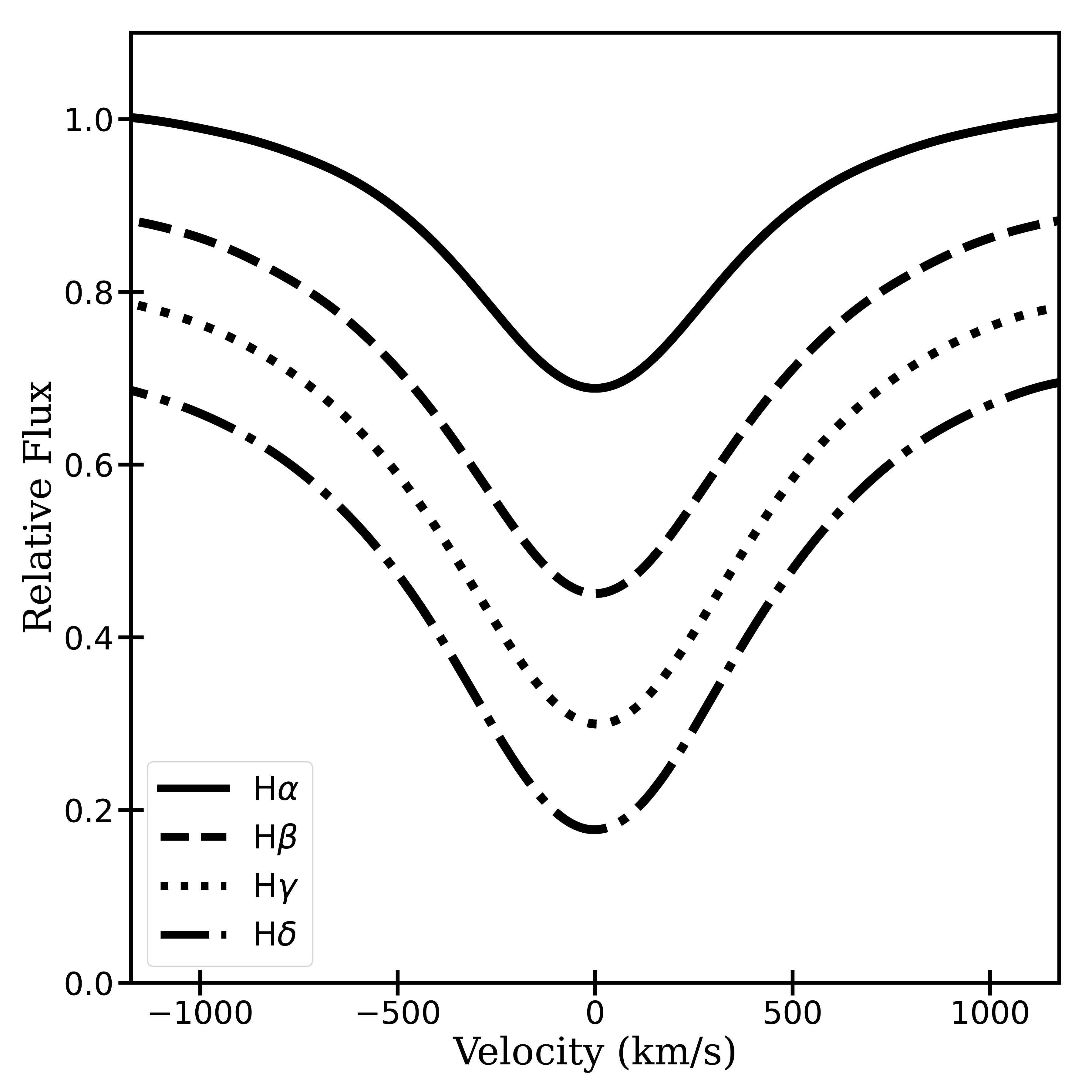}
\caption{Hydrogen Balmer series line profiles digitized from photographic observations shown in Figure \ref{fig:photographs_of_plates}. These line profiles are an average of the matching profiles across each of the four photographic plate spectra. The process used for combining the different exposures of the observed spectra allows for high signal-to-noise ratio, leaving relatively little noise in the line profiles. The flux of the H$\beta$, H$\gamma$ and H$\delta$ lines has been offset for ease of viewing.}
\label{fig:harvard_spectra_digital}
\end{figure}

We acquired four Balmer series spectra of Pleione during its last diskless phase (1906 to 1938 \citep{fro06, mcl38}) from the Harvard Astronomical Plate Collection. These spectra were recorded on photographic plates in 1927 at the Cambridge Observatory, and contain only photospheric flux unaffected by the presence of a circumstellar disk. These spectra were obtained using low-resolution dispersion objective prisms on a 24-inch Clark reflector, with a $0.6~\rm{m}$ aperture at a scale of $59.6~\rm{arcsec/mm}$. These photographic plates were previously unavailable in a digital format.

High-resolution photographs were taken of the emulsion side of plates designated as H02957, H02958, and H02959, according to classification system of the Harvard Astronomical Plate Collection. Figure~\ref{fig:photographs_of_plates} shows the averaged spectra acquired from each plate, with two spectra labelled (a) and (b) coming from plate H02957. Table \ref{tab:harvard_plates} provides the observational details for each plate.

The spectra were extracted from the photographs by the following method. After correcting for tilt on the plate, we averaged the logarithmic intensity at each wavelength across the vertical axis of the spectrum, effectively creating a time-averaged spectrum across the duration of the exposure. The spectra were wavelength-calibrated at the center of the H$\alpha$ and H$\delta$ lines, and sampled at every $1~\rm{\angstrom}$ from $7000~\rm{\angstrom}$ to $3990~\rm{\angstrom}$. The H$\epsilon$ line was not included in our data set due to low signal-to-noise ratio.  Figure \ref{fig:harvard_spectra_digital} shows the final H$\alpha$, H$\beta$, H$\gamma$ and H$\delta$ spectra, where each Balmer series line has been averaged across the four observations.

To acquire the relative fluxes from each photographic plate we required the non-linear response curve of each emulsion as a function of wavelength. Since the photographic plates also have other Pleiades cluster stars, we were able to extract the response curve from the star Atlas (HD 23850, 27 Tau), a B8III spectral-type star with no reports of H$\alpha$ variability in the literature over the past 100 years. Atlas lies near Pleione on the plane of the sky, and at a similar distance (387 ly for Pleione, 422 ly for Atlas, \citealt{van07}). The response curve was then reconstructed by equating Atlas' spectrum from each plate to a 2004~February spectrum acquired from the ELODIE archive \citep{mou04}. The spectra for Pleione were then extracted, and transformed as a function of wavelength according to the response curve. Figure \ref{fig:harvard_spectra_digital} shows our final H$\alpha$, H$\beta$, H$\gamma$ and H$\delta$ spectra which have been averaged across the four plates. We note that Atlas is brighter than Pleione ($3.5~\rm{mag}$ vs $5.0~\rm{mag}$ in the V-band, respectively, \citealt{duc02}). Therefore, for our method of obtaining the fluxes, we are assuming that both stars are appropriately exposed such that the main spectral features have photographic densities in the linear part of the response curve.

\subsection{Photometry} \label{subsec:photometry}

Pleione has been known to be photometrically variable since 1936 \citep{bin49}. V-band photometric observations from 1980 to 2010 were compiled from the following publications: \citet{sha88, hir76, hir77, hop80, hop82, dap81, boh84, boh85, boh86, boh88, sha92, sha97, pol11, tan07a}. Archival observations were acquired from the Hipparcos \citep{van07} mission and the ASAS-3 \citep{poj97} telescope archive.

Visible and infrared (IR) flux observations were obtained from the CDS Portal application from the Universit\'e de Strasbourg. The observations were compiled from catalogs which listed target objects within $0.5~\rm{arcsec}$ of Pleione's position. The catalogs included observations from 1990 to 2010, covering both the most recent Be and Be-shell phases. This data was sourced from \citet{sta07}, \citet{ega03} and the following missions: GAIA \citep{gaiaeDR3}, 2MASS \citep{skr06}, WISE \citep{wri10}, AKARI \citep{mur07}, IRAS \citep{neu84}, and Spitzer \citep{wer04}. These observations and the V-band photometry are presented for comparison against our precessing disk model in Section \ref{sec:results}.

Photographic magnitudes obtained during Pleione's last diskless epoch were also obtained. In 1918, \citet{par18} reported that Pleione had a photo-visual magnitude of $5.15~\rm{mag}$, and a M\"uller and Kempf visual magnitude of $5.08~\rm{mag}$. In 1922, \citet{lin22} also reported a photo-visual magnitude of $5.15~\rm{mag}$. These magnitudes were used along with UV spectra to constrain Pleione's stellar parameters in Section \ref{sec:stellarparameters}.

\subsection{Polarimetry} \label{subsec:polarimetry}

We acquired observations of the optical (BVRI) linear polarization from 2010 to 2021 using the IAGPOL polarimeter at the Pico dos Dias Observatory (OPD), operated by the National Astrophysical Laboratory of Brazil (LNA) in Minas Gerais, Brazil. These observations were reduced with packages developed by the B\textsc{eacon} group\footnote{http://beacon.iag.usp.br/}, and described in \citet{mag84, mag96} and \citet{car07}.

Archival polarization data of Pleione in the V-band were acquired from \citet{hir07}, who regularly observed the star from 1975 to 2004. Observations of the V-band polarization were also acquired from the archive for the Lyot Spectropolarimeter\footnote{http://www.sal.wisc.edu/PBO/LYOT/} and the Halfwave Spectropolarimeter (HPOL) at the University of Wisconsin-Madison Pine Bluff Observatory, which were reduced by \citet{dra14}. Additional polarimetric observations were extracted from \citet{hir07} using the \textit{WebPlotDigitizer} tool from \citet{roha20}. These archival data and our observations are presented in comparison to the percent polarization and polarization position angles of our precessing disk model in our results (Section \ref{sec:results}).

\begin{table}
	\centering
	\caption{The best-fitting stellar parameters for Pleione computed with \emcee.}
	\begin{tabular}{@{} l
                    @{\hspace*{4mm}}l
                    @{\hspace*{10mm}}l
                    @{\hspace*{4mm}}l
                    @{\hspace*{4mm}}l @{}}
		\hline
		\hline
		Best-Fit &  & & Derived &  \\
		Parameters & Values & & Parameters & Values \\
		\hline
		$M$ [$M_{\odot}$]  &  $4.1\substack{+0.2 \\ -0.2}$ & & $L$ [$L_{\odot}$]  &  $380\substack{+80 \\ -70}$ \\
		$W$  &  $0.8\substack{+0.1 \\ -0.1}$ & & $T_{\rm eff}$ [K]  &  $14000\substack{+100 \\ -100}$ \\
		$t/t_{\rm ms}$  &  $0.7\substack{+0.1 \\ -0.1}$ & & $\log g$  &  $4.0\substack{+0.1 \\ -0.1}$ \\
		$i$ [$^{\circ}$]  &  $61\substack{+9 \\ -8}$ & & $R_{\rm pole}$ [$R_{\odot}$] & $3.3\substack{+0.3 \\ -0.3}$ \\
		$d$ [pc]  &  $134\substack{+5 \\ -5}$ & & $R_{\rm eq}$ [$R_{\odot}$] & $4.4\substack{+0.6 \\ -0.5}$ \\
		$E(B-V)$  &  $0.09\substack{+0.02 \\ -0.02}$ & & $R_{\rm eq}$/$R_{\rm pole}$ & $1.3\substack{+0.1 \\ -0.1}$ \\
		\hline
		\hline
	\end{tabular}
	\label{tab:28taustellarpara}
\end{table}

\begin{table}
	\centering
		\caption{The mass, critical fraction of rotation and age used for the BeAtlas grid of models.}
	\begin{tabular}{lllll}
		\hline
		\hline
		Parameter & Grid Values & \\
		\hline
		$M$ [M$_{\odot}$]  &  $1.7, 2, 2.5, 3, 4, 5, 7, 9, 12, 15, 20$  &  \\ 
		$W$         &  $0.00, 0.33, 0.47, 0.57, 0.66, 0.74, 0.81, 0.93, 0.99$  &  \\ 
		$t/t_{\rm ms}$  &  $0, 0.25, 0.5, 0.75, 1, 1.01, 1.02$  &  \\  
		\hline
		\hline
	\end{tabular}
	\label{tab:beatlasgrid}
\end{table}

\begin{table}
	\centering
	\caption{Adopted stellar parameters used as priors in \emcee\ when fitting to the BeAtlas grid.}
	\begin{tabular}{lllll}
		\hline
		\hline
		Parameter & Value & Reference & \\
		\hline
		parallax [mas]  &  $7.24 \pm 0.13$  & \citet{gaiaeDR3} &  \\  
		$v$sin$(i)$ [km/s]  &  $286 \pm 16$  & \citet{fre05} &  \\  
		$i~[^\circ]$ &  $60 \pm 10$ & \citet{hir07} &  \\ 
		\hline
		\hline
	\end{tabular}
	\label{tab:emceeprior}
\end{table}

\begin{figure*}[ht]
	\centering
	\makebox[\textwidth][c]{\includegraphics[width = 1\textwidth]{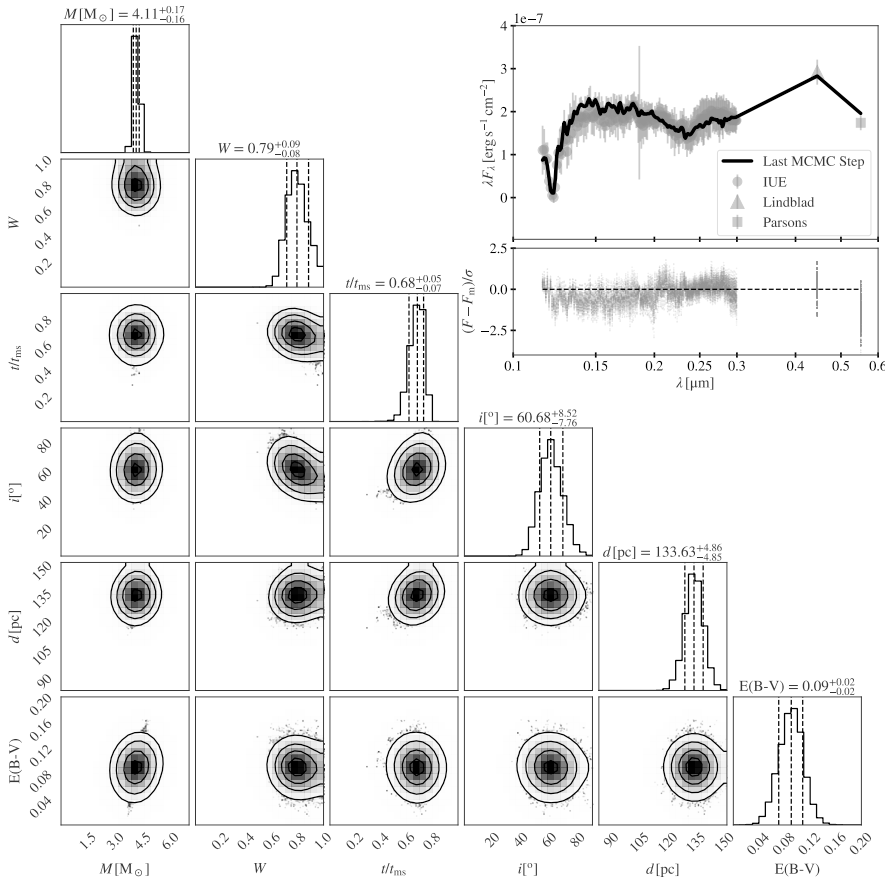}}
	\caption{Best-fitting stellar parameters based on UV and visible observations of Pleione. The probability density functions of each parameter are shown on the main diagonal axis while the intersection for each parameter shows the correlation map. The six parameters used in the fitting procedure are the stellar mass $M$, the rotation rate $W$, time of life on the main sequence $t/t_{ms}$, stellar inclination $i$, distance $d$, and interstellar reddening $E(B-V)$. The subfigure in the top-right corner shows the most probable fit model to the UV spectra and diskless visible magnitudes, and the residuals below.}
	\label{fig:bemcee}
\end{figure*}

\begin{figure}
\centering
\includegraphics[width=\columnwidth]{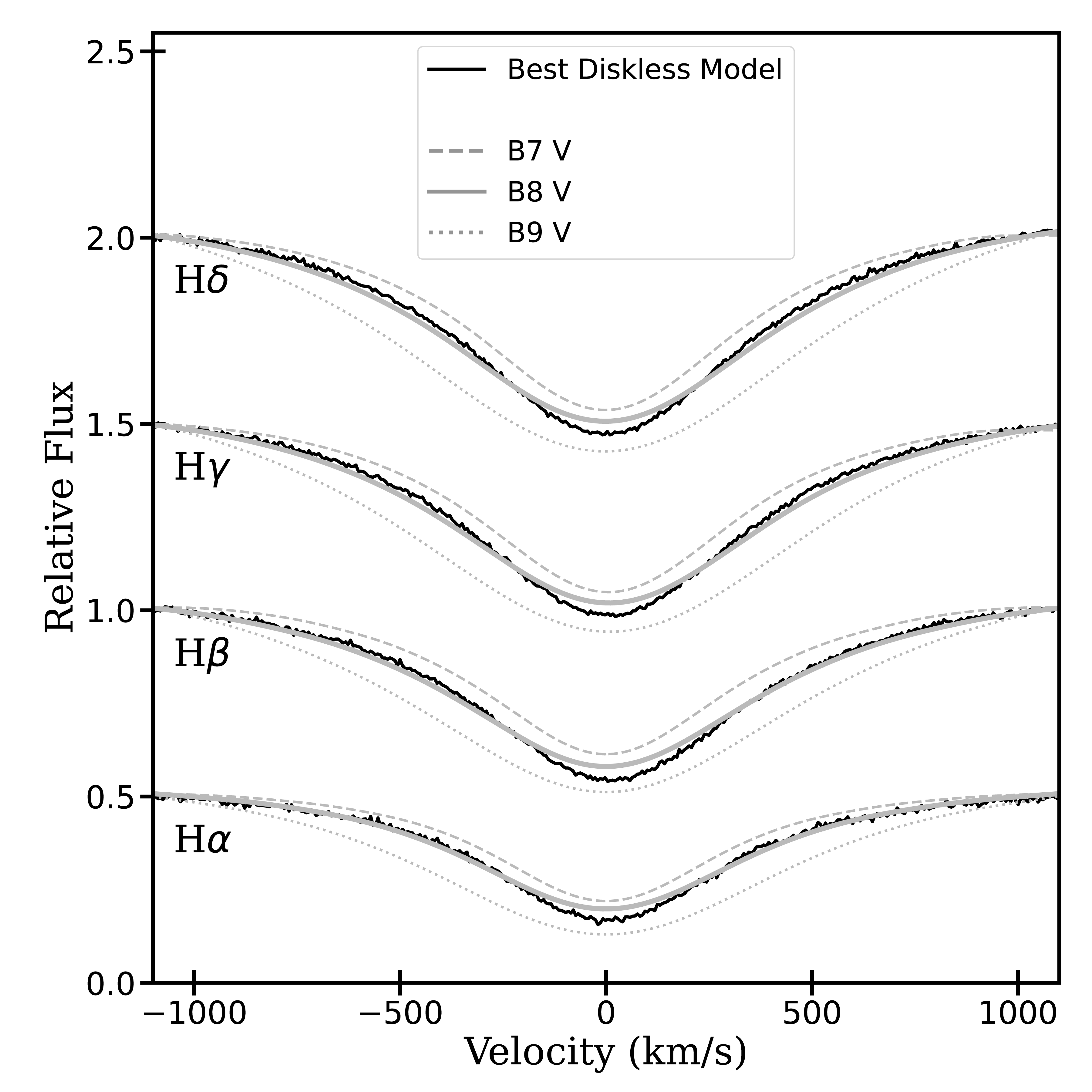}
\caption{Hydrogen Balmer series line profiles of standard B7 V, B8 V and B9 V stars computed with \hdust\ using stellar parameters from \citet{cox00} compared to the model lines computed using the stellar parameters in Table \ref{tab:28taustellarpara}.}
\label{fig:spectral_type_diskless}
\end{figure}

\begin{figure}
\centering
\includegraphics[width=\columnwidth]{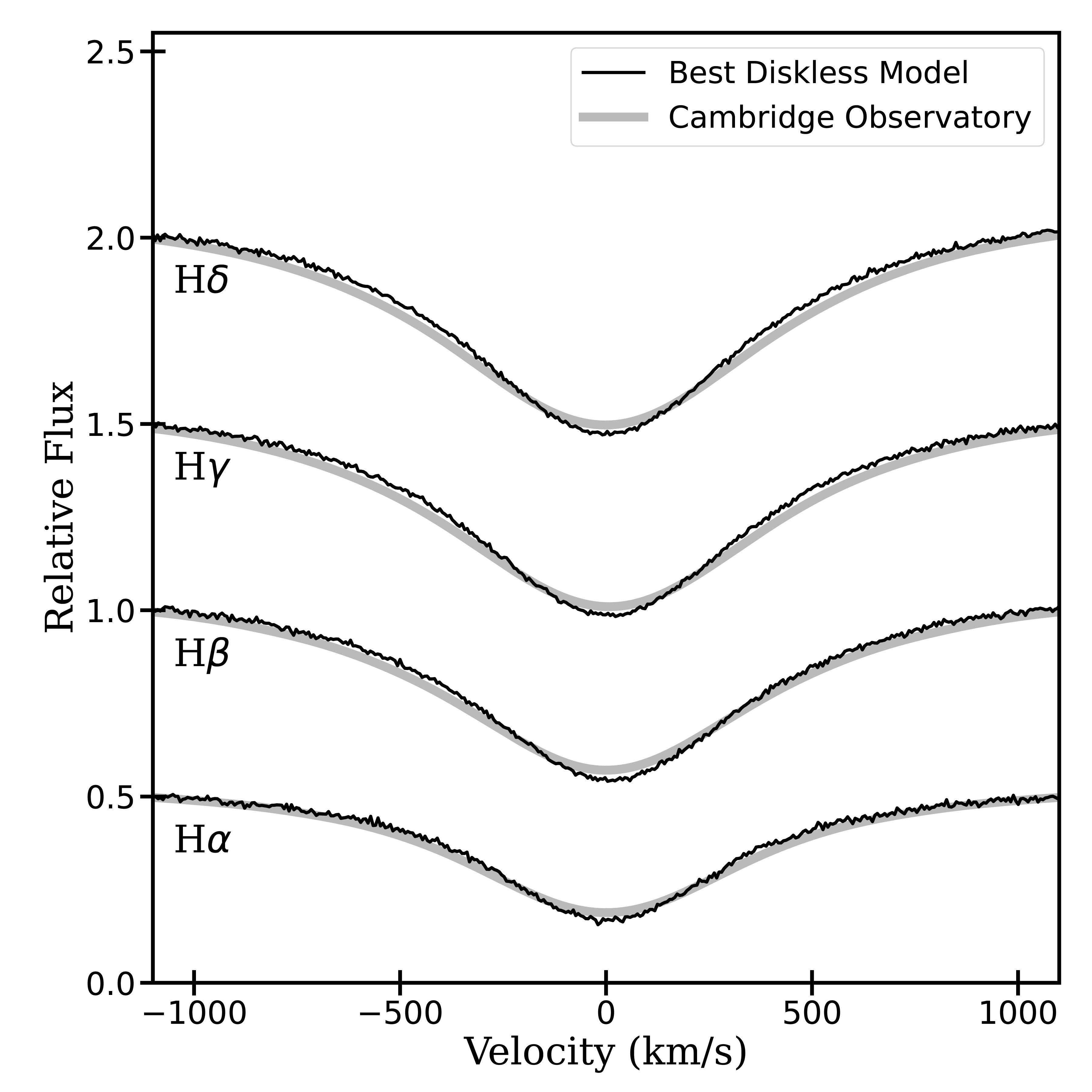}
\caption{A comparison of Pleione's H$\alpha$, H$\beta$, H$\gamma$, H$\delta$ line profiles from 1927~(Figure~\ref{fig:harvard_spectra_digital}), which correspond to the last diskless phase, with synthetic profiles computed using the best-fit stellar parameters determined with emcee and the BeAtlas grid~(Table~\ref{tab:28taustellarpara}).}
\label{fig:h02957_diskless_halpha}
\end{figure}

\newpage

\section{Stellar Parameters} \label{sec:stellarparameters}

We used IUE UV continuum observations and diskless photographic and photo-visual band photometry from \citet{par18} and \citet{lin22} to determine Pleione's stellar parameters using Monte Carlo methods, using the same procedure as \citet{mar21}. We determined the stellar mass $M$, critical fraction of rotation $W$ \citep[as defined in equation 6 of][]{riv13}, age $t/t_{\rm ms}$ (where $t_{\rm ms}$ is the main sequence lifetime, and is related to the fraction of hydrogen remaining in the stellar core), inclination $i$, distance $d$, and the degree of interstellar reddening E($B-V$). Additional parameters, listed in Table \ref{tab:28taustellarpara}, were self-consistently derived.

Overall, a grid of $770$ diskless Be star model spectra, called BeAtlas~\citep{mot19}, were fit to the observations. The models were computed using the 3D non-LTE radiative transfer code \hdust~\citep{car06}. \hdust\ computes model spectra from given stellar parameters, including $M$ and $W$, as well as the polar radius ($R_{\rm pole}$), luminosity ($L$) and gravity darkening exponent ($\beta$). The values of $R_{\rm pole}$, $L$ and $\beta$ are correlated to $M$ and $W$ through the stellar evolutionary models of \citet{geo13}. Table \ref{tab:beatlasgrid} lists the grid size and spacing. The excess flux due to interstellar reddening, E($B-V$), was also included as a free parameter, with its effect on the simulated spectrum being determined using the \citet{fit99} prescription.

We used the Markov Chain Monte Carlo (MCMC) routine \emcee\ by \citet{for13} to determine which models from the BeAtlas grid best reproduce the observed UV spectrum using \hdust. This MCMC routine was used to generate a list of stellar parameters inside pre-determined ranges that were weighted using literature values for parallax, $v\sin i$, and $i$ (Table \ref{tab:emceeprior}), and then the goodness of fit for each model was computed based on the fit of the resulting continuum to the observation using a $\rm{log}(\chi^2)$ likelihood function. The fitting procedure reached convergence using 30 walkers, and 50,000 steps, with a burn-in of 5,000 steps, which were chosen following the guidelines of \citet{for13}. Further details on the \emcee\ fitting process are given in \citet{mot19}.

We quantified the impact of the variable UV flux on the stellar parameters by separately fitting the low (1979~July) and high (1995~March) extremes~(recall Figure~\ref{fig:IUE_spectra}), along with the diskless visible photometry. We found $M$ in the range $3.98$ to $4.23~\rm{M_{\odot}}$ and $E(B-V)$ from $0.07$ to $0.10~\rm{mag}$, while the other parameters did not change. By averaging the UV observations we find a set of stellar parameters which reproduce the UV and visible spectrum with $\chi_{\nu}^2 = 1.65$. The errors on these parameters include the ranges of $M$ and $E(B-V)$ from the UV extremes.

Figure \ref{fig:bemcee} depicts the stellar parameters when fitting to the averaged UV spectra and diskless visible photometry. The probability density functions (PDF) for each stellar parameter are shown along the main diagonal as histograms, and the intersections of the parameters show the corresponding correlation maps. The width of the PDF indicates how well the parameter is constrained, while the center dashed line on each histogram shows the most probable value, and the dashed lines on the left and right of center indicate the first and third quartiles. In the upper right corner of Figure~\ref{fig:bemcee} the SED is shown along with the best-fit model. Below the SED, the residuals between the observations and model are shown. The best-fit stellar parameters are summarized in Table \ref{tab:28taustellarpara}.

As a final check of our best-fit stellar characteristics, we compared the model stellar absorption profiles to what would be expected for late B-type stars, since in the literature the spectral type of B8 V is often applied to Pleione \citep[e.g.,][]{gul77, hel78}. Specifically, we compared the synthetic hydrogen Balmer series spectra to model profiles calculated with \hdust\ based on standard B7 V, B8 V and B9 V star parameters from \citet{cox00}. Figure \ref{fig:spectral_type_diskless} shows our best-fit diskless model most closely resembles the B8 V hydrogen line profiles. The stellar parameters were further verified through comparison with the diskless Balmer series lines, previously described in Section \ref{subsubsec:digitizationofdisklessspectra} and Figure~\ref{fig:h02957_diskless_halpha} shows the computed lines of the best-fit model overlaid with the digitized observations. We find the H$\alpha$, H$\beta$, H$\gamma$, and H$\delta$ fit the observations with $\chi_\nu^2 = 1.27$, $1.46$, $1.58$, $1.53$, respectively.


\section{Disk Modelling} \label{sec:results}
\subsection{A Precessing Disk Model} \label{subsec:results_1}

\begin{figure}
\centering
\includegraphics[width=\columnwidth]{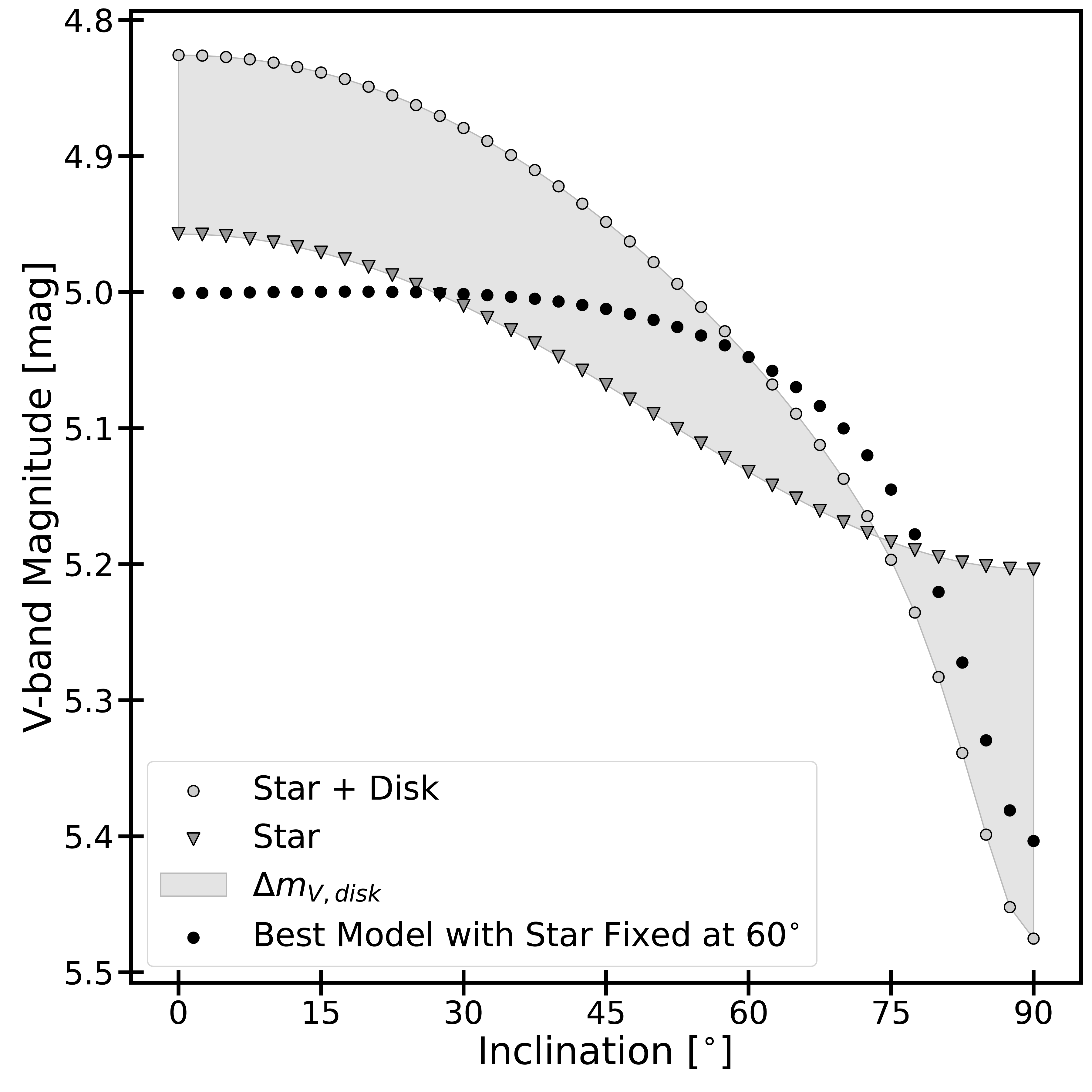}
\caption{Change in the observed V-band apparent magnitude for different system inclinations, using the models with the best-fit stellar and disk parameters. Shown are a diskless star at different inclinations (triangles), a star and untilted disk which incline together (open circles), and a star fixed at $60\rm{^{\circ}}$ while the disk inclination changes (filled circles).}
\label{fig:vband_vs_inc}
\end{figure}

\begin{figure*}
    \centering
    \makebox[\textwidth][c]{\includegraphics[width = 1\textwidth]{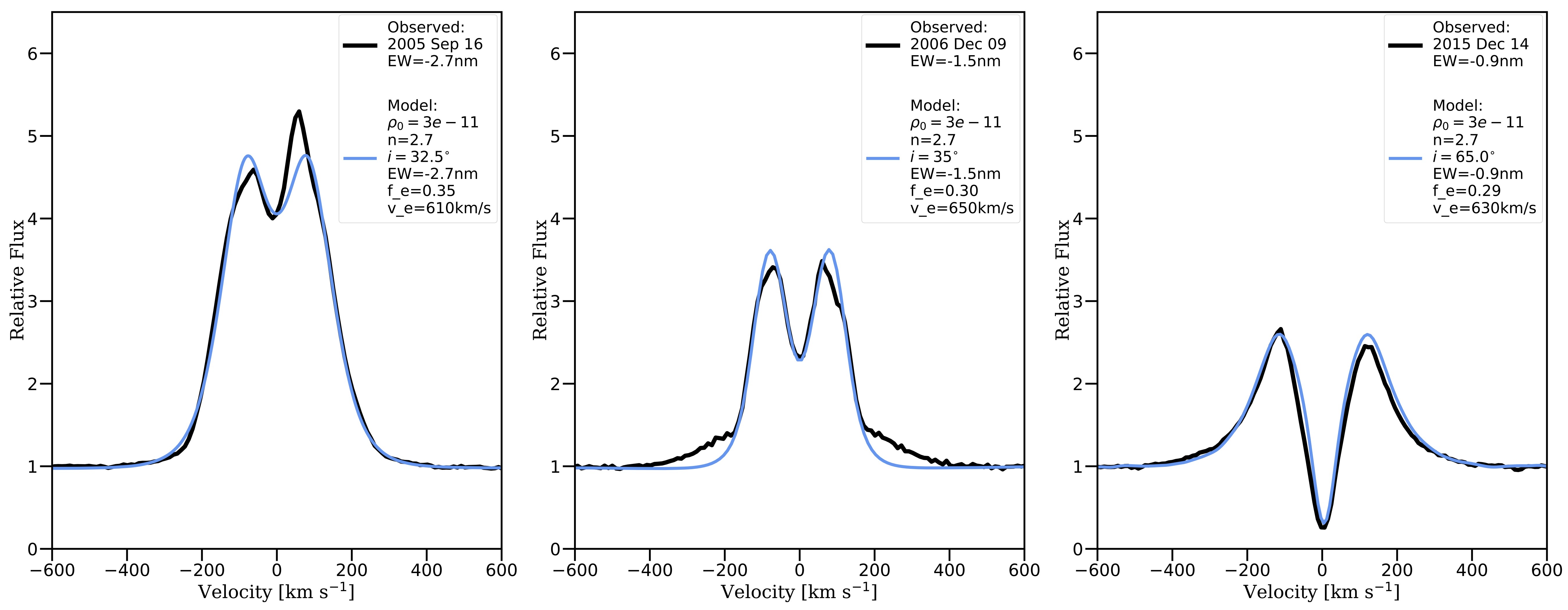}}
    \caption{A comparison of the 2005~September~16 (left), 2006~December~9 (center) and 2015~December~14 (right) H$\alpha$ observations from Lowell Observatory (black) to the best-fit model spectra (blue). The inclinations for the model spectra and convolving parameters are given in the legend.}
    \label{fig:halpha_fit_all_phases}
\end{figure*}

We computed a grid of 94,720 Be star and disk models using \hdust, to model the changes observed in the H$\alpha$ emission. This grid consists of axisymmetric, non-isothermal, thin-disk envelopes. The vertical disk density distributions are Gaussian, while the radial distributions follow a power-law of the form $\rho (r) \propto \rho_0 r^{-n}$, where $\rho_0$ is the density at the base of the disk ($r = R_{eq}$), and $n$ is the density falloff exponent. Values of $\rho_0$ were explored in the range of $1 \times 10^{-13}$ to $1 \times 10^{-10}~\rm{g~cm^{-3}}$ for every tenth of a magnitude, and the density falloff exponent $n$ over the range of $2.0$ to $3.5$ in steps of $0.1$. Each model was computed with the outer disk radius $R_{\rm out}$ at 5, 10, 15, 20, 25, 50, 75 and 100 stellar equatorial radii ($R_{\rm eq}$), and using inclinations from $0$ to $90\rm{^\circ}$ in steps of $2.5\rm{^\circ}$.

The line broadening effect due to non-coherent electron scattering within the disk \citep{aue68, poe79b} was approximated as it is not accounted for in \hdust\ simulations. For each synthetic profile, a fraction ($F_{\rm e}$) of the synthetic H$\alpha$ flux (where $F_{\rm e}$ ranged from $0.2$ to $0.4$) was convolved with a Gaussian of width equal to the electron velocity $v_e$~(where $v_e$ ranged from $400$ to $800~\rm{km~s^{-1}}$). The remaining $1 - F_e$ of the flux was left unbroadened. Through MCMC fitting with \emcee, $F_{\rm e}$ and $v_e$ were treated as free parameters within the given ranges, to determine the best possible fit for each model to the observed H$\alpha$ spectra~\citep[see][for more details on this procedure]{mar21}.

\begin{table}
	\centering
		\caption{Best-fit disk parameters obtained for each H$\alpha$ line emission profile.}
	\begin{tabular}{cccccc}
		\hline
		\hline
		Date   & MJD & $i$ [$^{\circ}$] & $R_{\rm{out}}$  & $F_e$    & $v_e$ \\
		  & (+2400000.5) &  & $\rm{[R_{eq}]}$ &  & $\rm{[{km~s^{-1}}]}$ \\
		\hline
	
        2005 Aug 22 & 53604 & 30.0 & 15 & 0.34 & 610  \\
        2005 Sep 16 & 53629 & 32.5 & 15 & 0.35 & 610  \\
        2005 Sep 17 & 53630 & 32.5 & 15 & 0.35 & 610  \\
        2005 Oct 11 & 53654 & 32.5 & 15 & 0.35 & 610  \\
        2005 Nov 18 & 53692 & 32.5 & 15 & 0.35 & 610  \\
        2005 Dec 20 & 53724 & 32.5 & 15 & 0.35 & 610  \\
        2006 Jan 14 & 53749 & 32.5 & 15 & 0.35 & 610  \\
        2006 Jan 24 & 53759 & 32.5 & 15 & 0.35 & 610  \\
        2006 Feb 07 & 53773 & 32.5 & 15 & 0.35 & 610  \\
        2006 Mar 17 & 53811 & 32.5 & 15 & 0.35 & 610  \\
        2006 Dec 09 & 54078 & 35.0 & 15 & 0.30 & 650  \\
        2007 Jan 26 & 54126 & 35.0 & 15 & 0.33 & 650  \\
        2007 Feb 03 & 54134 & 35.0 & 15 & 0.31 & 650  \\
        2007 Dec 18 & 54452 & 37.5 & 15 & 0.25 & 640  \\
        2008 Nov 14 & 54784 & 40.0 & 15 & 0.25 & 640  \\
        2008 Nov 15 & 54785 & 40.0 & 15 & 0.24 & 640  \\
        2008 Nov 15 & 54785 & 40.0 & 15 & 0.24 & 640  \\
        2009 Dec 06 & 54994 & 42.5 & 15 & 0.28 & 640  \\
        2013 Dec 10 & 56636 & 60.0 & 15 & 0.30 & 630  \\
        2013 Dec 10 & 56636 & 60.0 & 15 & 0.30 & 630  \\
        2013 Dec 11 & 56637 & 60.0 & 15 & 0.30 & 630  \\
        2015 Apr 01 & 57113 & 65.0 & 15 & 0.29 & 630  \\
        2015 Apr 01 & 57113 & 65.0 & 15 & 0.29 & 630  \\
        2015 Dec 13 & 57369 & 65.0 & 15 & 0.29 & 630  \\
        2015 Dec 14 & 57370 & 65.0 & 15 & 0.29 & 630  \\
        2017 Dec 20 & 58107 & 75.0 & 15 & 0.31 & 610  \\
        2017 Dec 20 & 58107 & 75.0 & 15 & 0.31 & 610  \\
        2017 Dec 20 & 58107 & 75.0 & 15 & 0.31 & 610  \\
        2017 Dec 21 & 58108 & 75.0 & 15 & 0.31 & 610  \\
        2018 Dec 19 & 58471 & 77.5 & 15 & 0.33 & 610  \\
        2018 Dec 20 & 58472 & 77.5 & 15 & 0.33 & 610  \\
        2019 Dec 18 & 58835 & 80.0 & 15 & 0.35 & 630  \\
        2019 Dec 18 & 58835 & 80.0 & 15 & 0.35 & 630  \\
        2019 Dec 19 & 58836 & 80.0 & 15 & 0.35 & 630  \\
        2019 Dec 19 & 58836 & 80.0 & 15 & 0.35 & 630  \\
		\hline
		\hline
	\end{tabular}
	\label{tab:28taudiskpara}
\end{table}

\begin{figure}
\centering
\includegraphics[width=\columnwidth]{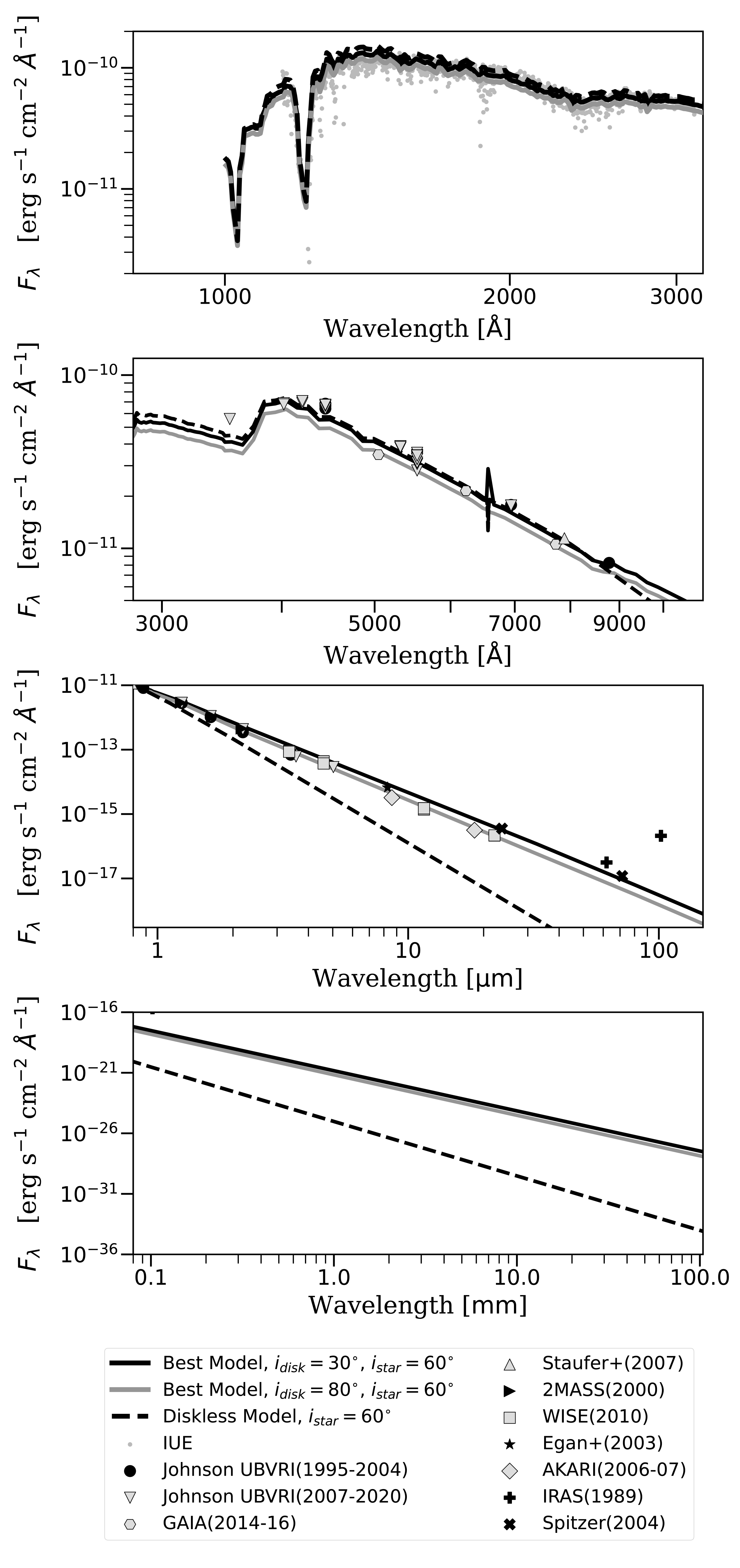}
\caption{SED of Pleione showing the best-fit disk and diskless models in comparison to observations in the UV (top), visible (upper middle), IR (lower middle), and radio (bottom). The observations have been separated into those taken during the Be phase (black) and during the Be-shell phase (grey). The IUE observations from the Be phase are in dark grey for ease of viewing. The radio SED is shown for the best-fit disk and diskless models despite no radio observations being available for comparison.}
\label{fig:SED_fit}
\end{figure}

We find that the observed trend of the H$\alpha$ profiles from 2005 to 2019 (recall Figure \ref{fig:tycner_halpha_observations}) can be reproduced by a single disk model by varying the inclination. Our models were limited to axisymmetric disks in the stellar equatorial plane, which when viewed at different inclinations require the star to be inclined at the same angle as the disk. To account for tilting, the absolute flux of the diskless model at the appropriate $i$ was subtracted from the star and disk model at the same $i$ to obtain flux for the inclined disk. The inclined disk flux was then added to the flux of the diskless model at $i_{\rm star} = 60\rm{^{\circ}}$. We believe this first-order correction captures the most significant effect on the change in flux by the tilting of the disk. This correction was applied across the entire spectrum, and affected models at low $i$ values most. For example, at a disk inclination of $i = 0\rm{^{\circ}}$, correcting for the stellar component to be at $i = 60\rm{^{\circ}}$ (instead of $i = 0\rm{^{\circ}}$) caused an increase in normalized H$\alpha$ peak flux of less than $40\%$ for all models, and at $i = 30\rm{^{\circ}}$ the effect was less than $20\%$. In the V-band, this correction decreased the continuum level by $\sim0.2~\rm{mag}$ at $i = 0\rm{^{\circ}}$ models, $\sim0.15~\rm{mag}$ at $i = 30\rm{^{\circ}}$, and increased by $\sim0.05$ at $i = 75\rm{^{\circ}}$ (see Figure~\ref{fig:vband_vs_inc}).

Our best-fit model to the H$\alpha$ observations has density of $\rho_0 = 3 \times 10^{-11}~\rm{g~cm^{-3}}$ and $n = 2.7$, an outer disk radius of $R_{\rm out} = 100~R_{\rm eq}$, and a range of inclinations from $30^{\circ}$ to $80^{\circ}$. This single density model suggests that the mass-loss rate from the star was constant during this period. We note that for the models with radii of $15~R_{\rm eq}$ and larger, the improvements to the $\chi_\nu^2$ fit were not statistically significant. Therefore, we quote a lower limit of $15~R_{\rm eq}$ for the H$\alpha$ emitting region, while its real size remains undetermined.

Figure~\ref{fig:halpha_fit_all_phases} shows a sample of three H$\alpha$ profiles, chosen to illustrate how well the best-fit models match the data. The other fits to the observations that are not shown were of similar quality. The non-coherent electron scattering in the best-fit models has values of $F_e = 0.33 \pm 0.04$ and $v_e = 635 \pm 25~\rm{km~s^{-1}}$. This electron velocity corresponds to a temperature of $\sim8900~\rm{K}$, which is similar to the disk temperature of $\sim9000~\rm{K}$ at the outer edge of the H$\alpha$ emitting region given by \hdust\ models, and the isothermal disk temperature of $\sim8300~\rm{K}$ ($60\%$ of $T_{eff}$, \citealt{car06}). Our method of accounting for the broadening due to electron scattering more strongly affects the width of H$\alpha$ profiles formed at lower disk inclinations. As a result, profiles at different inclinations can have the same width, while having different strengths. The models fit with reduced $\chi^2$ values ranging from $1.33$ to $3.82$. Table \ref{tab:28taudiskpara} summarizes the best-fit model disk parameters obtained from each H$\alpha$ emission line observation.

During the recent Be to Be-shell transition, Pleione exhibited a rapid drop in the visible and IR continuum flux \citep{tan07a}. In Figure~\ref{fig:SED_fit}, we show that our best-fit model reproduces this drop in IR flux when considering the minimum ($30^{\circ}$) and maximum ($80^{\circ}$) inclinations found. The best-fit model to the 2005~September~16 H$\alpha$ profile fits the Be phase SED with reduced $\chi^2 = 1.93$, while the best-fit model to the 2015~December~14 observation fits the Be-shell SED with a reduced $\chi^2 = 1.65$. The UV continuum changed by less than 2\% between the phases~(i.e., within the error ranges). The predicted radio spectra of the simulations are shown for reference, however no radio continuum observations were available for comparison.

\begin{figure}
\centering
\includegraphics[width=\columnwidth]{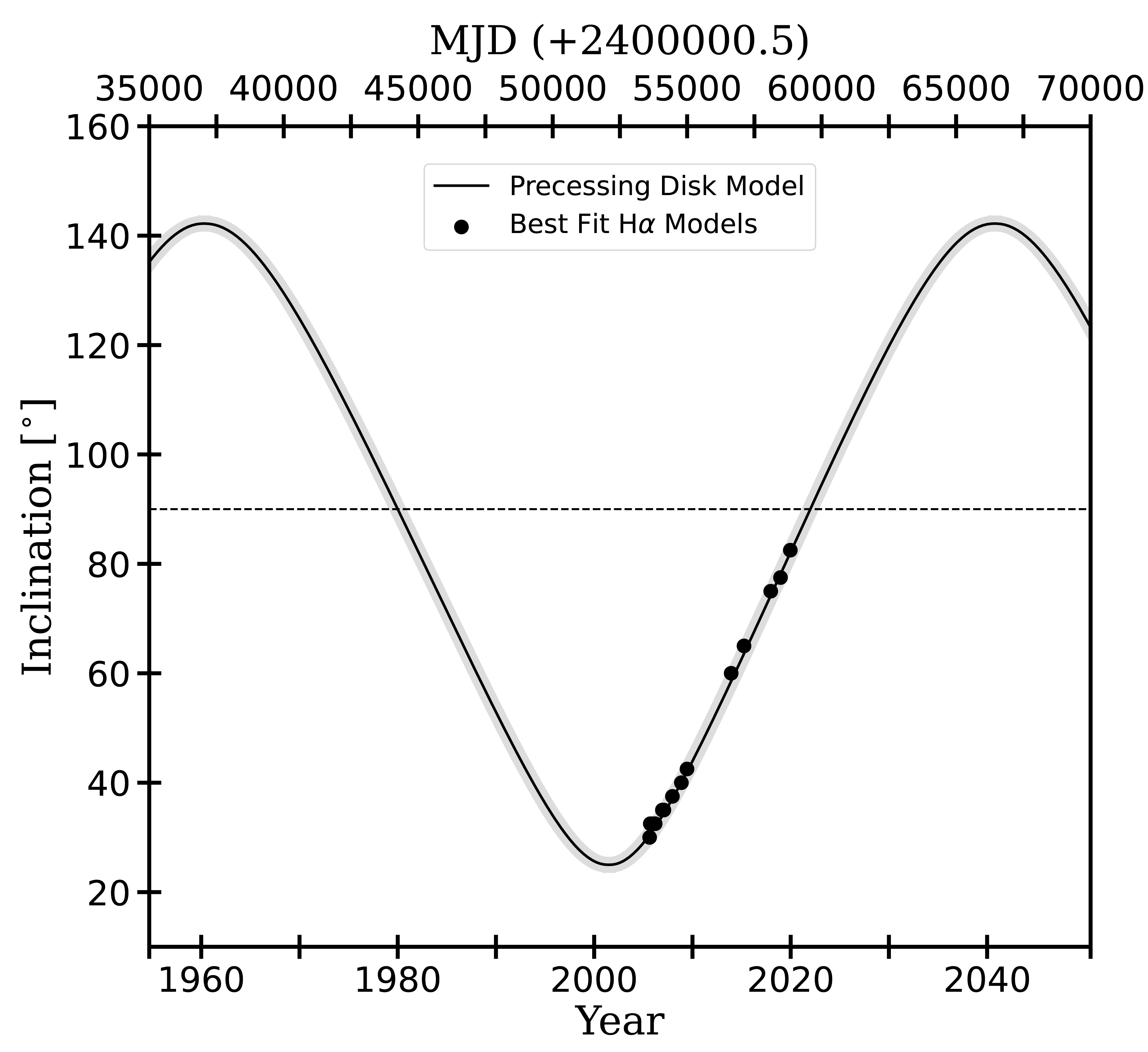}
\caption{Change in the stellar inclination with time determined from our best-fit models to the H$\alpha$ observations, along with our precessing disk model. The horizontal dashed line indicates $90^{\circ}$ inclination. Inclinations greater than $90^{\circ}$ occur when viewing the lower part of the disk.}
\label{fig:inc_vs_time}
\end{figure}

The inclinations of our best-fit model are shown in Figure \ref{fig:inc_vs_time} at the time of each observation. We find that over the course of our observations the disk's inclination changes at a rate of $\sim3.7\rm{^{\circ}}$ per year, and was once again viewed edge-on in 2021~November.

To investigate the scenario of a precessing disk as proposed by \citet{hir07}, we created a disk model by adopting equation 1 of \citet{dun06}, which describes the motion of a precessing galactic disk. This equation determines the inclination of the disk when provided with two values: $\gamma$, the angle between the precession axis and the observer's line of sight, and $\delta$, the angle between the disk's normal and the precession axis. We used MCMC fitting with \emcee\ to determine the values of $\gamma$ and $\delta$ that best reproduced the inclinations determined through H$\alpha$ fitting. We note that $\gamma$ is not fixed to the stellar spin-axis. This is a reasonable assumption since the dynamical simulations of \citet{mart11} show that the precession axis of a misaligned circumstellar disk lies normal to the orbital plane of the companion and is not coupled to the stellar spin axis. Since we do not know the value of the orbital inclination for Pleione, it is reasonable to regard the precession axis as a free parameter. The timing of the minimum inclination and the period of the precession were also fitted parameters. We find the best-fit precession model has $\gamma = 84\pm3^{\circ}$ and $\delta = 59\pm3^{\circ}$, with the minimum inclination of $25^{\circ}$ occurring in 2001~May, the maximum inclination of $144^{\circ}$ in 2040~September, and the precession period of $29400\pm100$ days, or $\sim80.5$ years. This precessing disk model is shown in Figure \ref{fig:inc_vs_time}, with the grey band indicating the uncertainty. However, since only a portion of the precession period is sampled by our observed H$\alpha$ profiles, the errors we find through this MCMC analysis are likely a lower bound.

\begin{figure}
\centering
\includegraphics[width=\columnwidth]{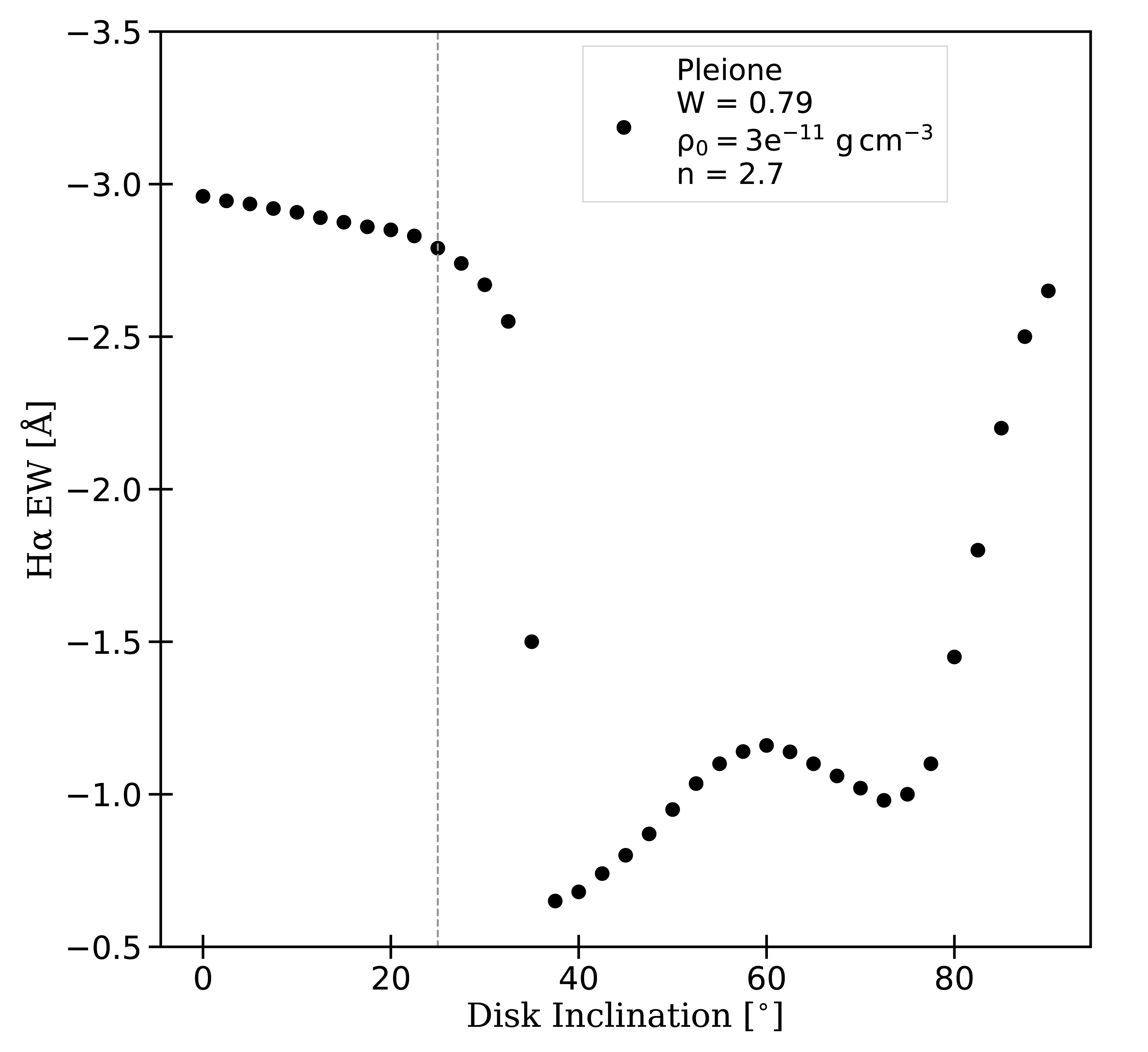}
\caption{Variation in the EW of the H$\alpha$ line profile with disk inclination, with the star fixed at $60^{\circ}$ for our precessing disk model. All predicted inclinations from $0^{\circ}$ to $90^{\circ}$ are shown, however the lower limit of the inclination range determined from H$\alpha$ profile fitting is indicated by the vertical dashed line.}
\label{fig:ew_vs_inc}
\end{figure}

\begin{figure}
\centering
\includegraphics[width=\columnwidth]{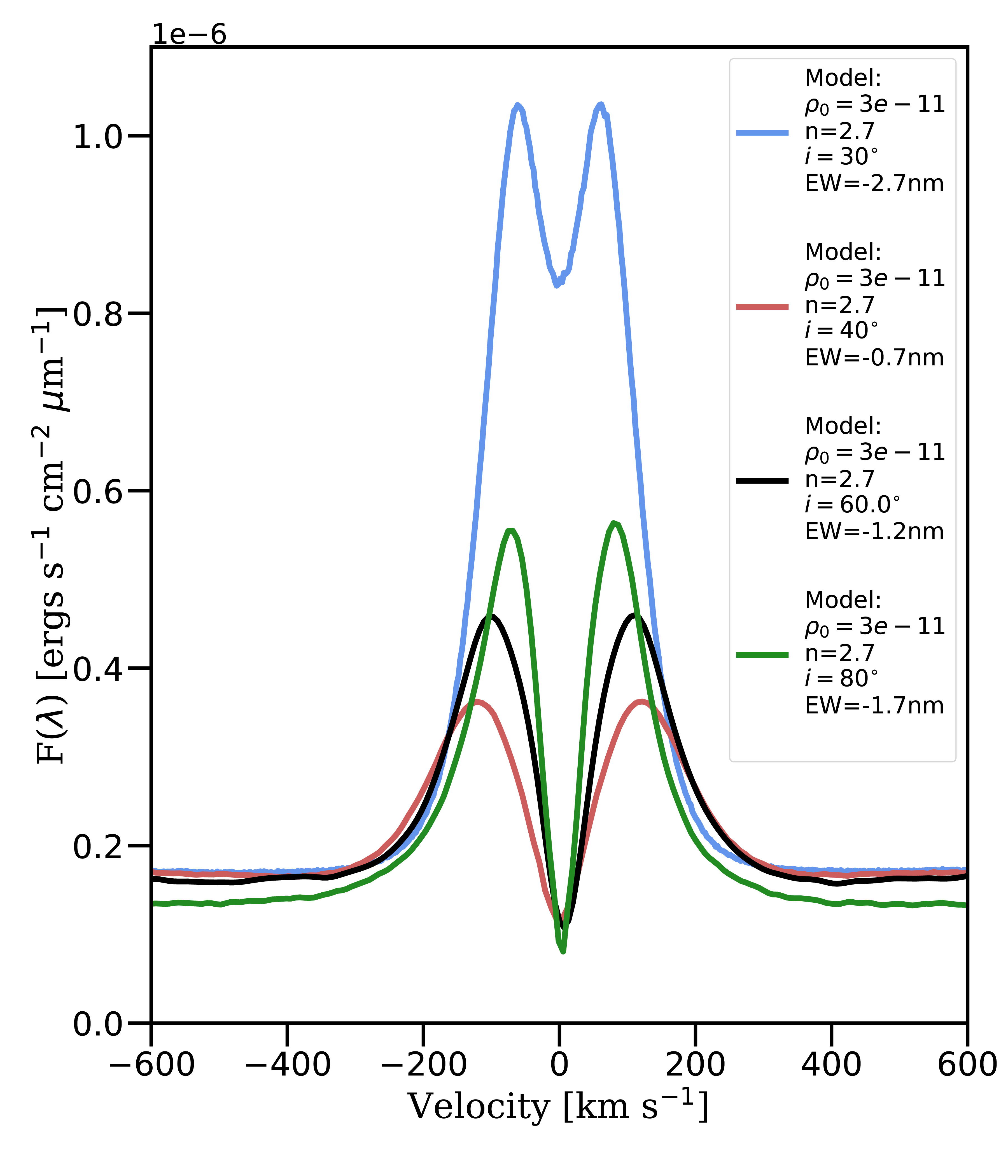}
\caption{Subset of the best-fit predicted H$\alpha$ profiles in absolute flux units, before convolution. The stellar inclination is fixed at $60^{\circ}$. The disk inclinations for each model spectrum are given in the legend.}
\label{fig:halpha_notnorm_notconv}
\end{figure}

The variation in H$\alpha$ EW with disk inclination is shown for the precessing disk model in Figure \ref{fig:ew_vs_inc}. Here, the H$\alpha$ EW is greatest at $-2.9~\rm{\AA}$ when the disk inclination is $0^{\circ}$. A secondary peak at $-2.6~\rm{\AA}$ occurs around $90^{\circ}$. Dramatic changes to the EW occur around particular inclinations. From $35^{\circ}$ to $40^{\circ}$, the EW decreases from $-2.5~\rm{\AA}$ to $-0.6~\rm{\AA}$ as the line transitions to a Be-shell profile. Beyond $40^{\circ}$, the line has a shell profile and the peak flux increases with the inclination while also becoming narrower. At higher inclinations as the disk becomes edge-on, the stellar continuum emission falls more rapidly, resulting in increases in our normalized H$\alpha$ spectra and the corresponding EW. This effect is illustrated in Figure \ref{fig:halpha_notnorm_notconv}. The Figure shows the H$\alpha$ spectrum for different inclinations with the star fixed at $60^{\circ}$, prior to normalization and the convolution process. Here, we can see the continuum flux around H$\alpha$ in absolute units decreases more rapidly at high disk inclinations.

\begin{figure*}
\centering
\includegraphics[width=\textwidth]{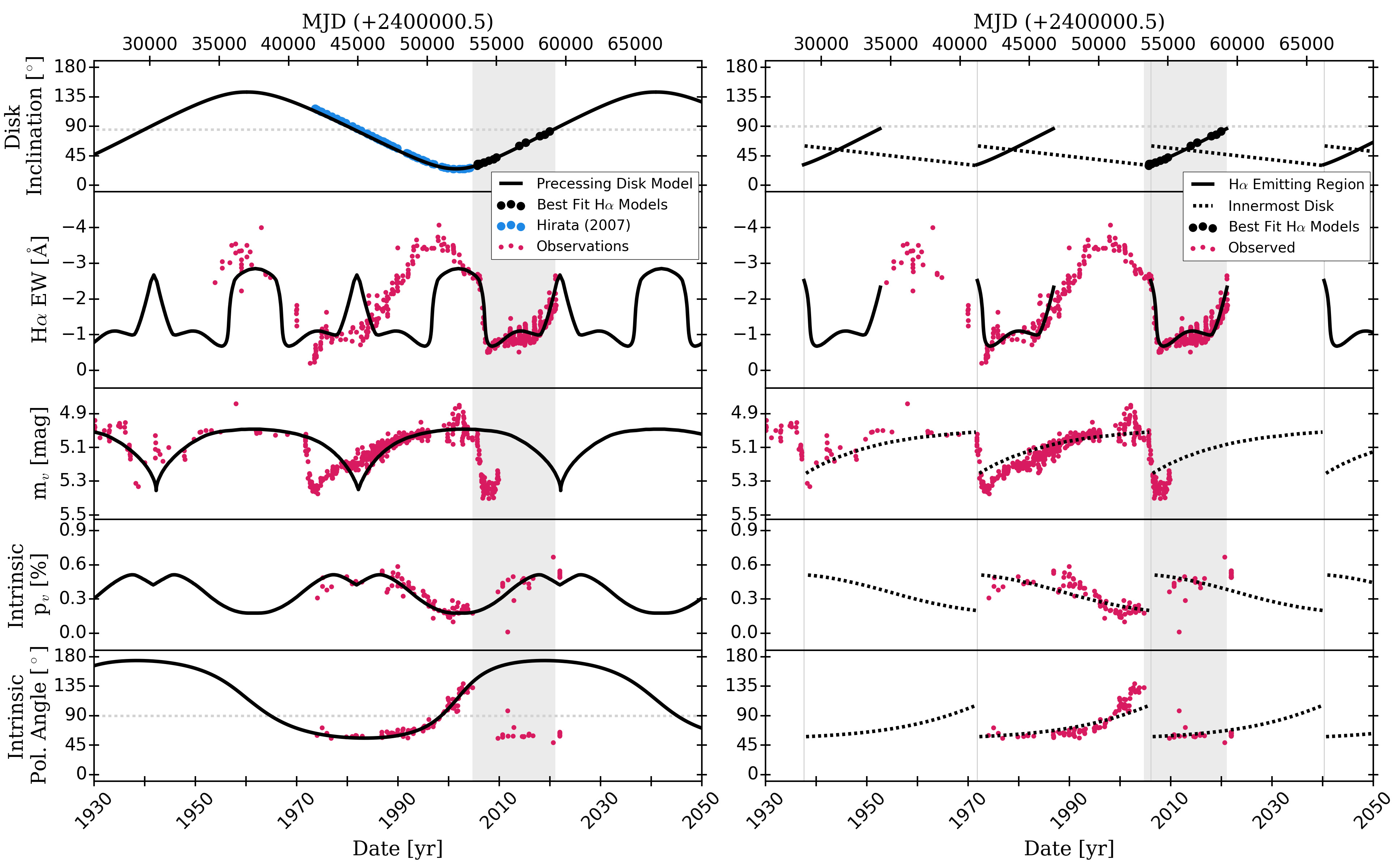}
\caption{\textit{Left:} Comparison of the precessing disk model (black, solid line) with an $80.5$ year period to archival observations (red), and to the inclinations from \citet{hir07}'s model (blue) and from our best-fit H$\alpha$ model (black). The light grey vertical band indicates the region over which our H$\alpha$ observations were modelled. In the top panel, the grey horizontal dashed line shows the inclination of $83.6^{\circ}$ that the precession is centered about. \textit{Right:} The same as the left side but instead showing the ad-hoc disk tearing disk model. The precessing H$\alpha$ emitting region (black, solid line) precesses for the first $15$ years of the $34$ year Be-shell to Be phase cycle. In the same period, the innermost disk (black, dotted line) gradually transitions from the stellar equator at $60^{\circ}$ to $30^{\circ}$. The thin, vertical grey lines indicate each disk tearing event. \citet{hir07}'s inclinations are not included as they follow the precession of a disk in V-band polarization, not H$\alpha$ emission.}
\label{fig:plotall_1}
\end{figure*}

We extrapolated the precessing disk model inclinations over 120 years, from 1930 to 2050. The left side of Figure \ref{fig:plotall_1} shows the inclinations of the extrapolated model, along with the H$\alpha$ EW, V-band photometry, V-band polarization level, and the V-band polarization position angle of our best-fit model at the inclinations of the precessing disk model (the right side of this Figure is discussed in Section \ref{subsec:results_2}). The observations previously described in Section \ref{sec:observations} are overplotted for comparison, as well as the inclinations that \citet{hir07} derived from their model. As our disk models are axisymmetric, the observables produced are symmetric about $90\rm{^{\circ}}$.

\begin{figure}
\centering
\includegraphics[width=\columnwidth]{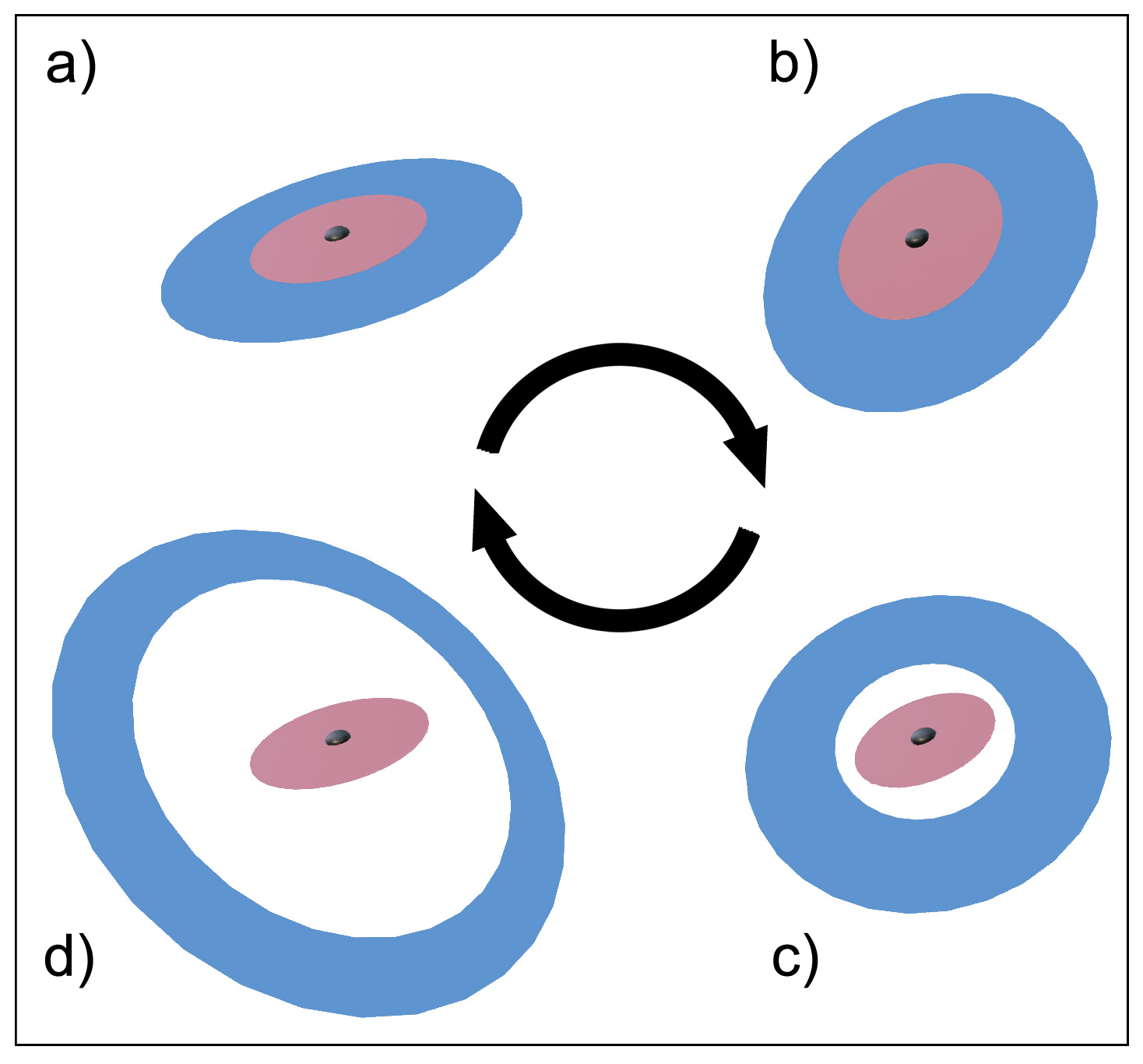}
\caption{Schematic of the disk tilting and tearing model. The red region of the disk corresponds to the innermost disk, and the blue region corresponds to the outer region which eventually separates. Stage one of our model is represented by a) and b), where the disk is whole and becomes progressively more tilted away from the equator from a) to b). Stage two is represented by c) and d), where the outer region tears, precesses, and eventually dissipates. In stage two, the innermost disk gradually returns to the original inclination at the stellar equator while continuing to build with constant mass injection prior to the cycle repeating. The sizes of the coloured sections are for illustrative purposes only and do not represent actual dimensions.}
\label{fig:schematic}
\end{figure}

In the second panel on the left side of Figure \ref{fig:plotall_1}, we show the trend of the H$\alpha$ EW of our precessing disk model, which is limited to the inclinations determined from the line profile fitting. As the inclination range is $>25^{\circ}$, the maximum EW observed is $-2.8~\rm{\AA}$. Note that in this Figure the grey shaded region is the same as what is shown in Figure \ref{fig:ew_vs_inc} to the right of the vertical dashed line. Overall, the trends in the H$\alpha$ EW, are mirrored about $90^{\circ}$ due to the symmetry of the precessing disk.

From 2005 to 2019, the best-fit tilted disk model fits our H$\alpha$ observations with reduced $\chi^2 = 1.6$. Outside of this period, our models cannot reproduce the observed trend using the $\sim80.5$ year precession period. We note that while a shorter period would improve the timing of the fit to the previous Be-shell phase in the 1970's, it also would make the model progress through its H$\alpha$ EW trend faster than what is observed.

In the third panel on the left side of Figure \ref{fig:plotall_1}, we see our precessing disk model is able to reproduce the gradually increasing V-band magnitude, from 1988 to 2004. In 2004, the model reaches an average observed magnitude of $\sim5.02~\rm{mag}$, before the model and observations diverge in 2007 when the magnitude was observed to rapidly drop by $0.3~\rm{mag}$. During this time, our precessing disk model takes $\sim14$ years to decline to this value. The minimum brightness of our model is $\sim5.38~\rm{mag}$, which occurs in 2021. As the rapid Be to Be-shell transition occurs every $\sim34$ years while half of the precession period is $\sim40$ years, we see that the previous minimum in brightness occurs in 1982, just $\sim8$ years after the observed minimum.

The polarization level of our best-fit model is shown in the fourth panel of the left side of Figure \ref{fig:plotall_1}. The maximum polarization level of $\sim0.53\rm{\%}$ is observed at $\sim70^{\circ}$. Projecting our best-fit models forward in time along inclination curve produces the polarization signature of a tilted disk, similar to those shown in \citet{mar18}. The polarization from the model fits the observations with reduced $\chi^2 = 1.7$.

To get the polarization position angle of our precessing disk model, we used equation 2 from \citet{dun06} to first extract the angle between the precession axis of the star projected onto the sky and the normal of the disk projected onto the sky. This angle has a constant offset from the polarization position angle, which is the angle between north on the sky and the projected precession axis of the star. We find this angle is $\sim115^{\circ}$ to align our best-fit precession model with the observed, intrinsic polarization position angle. The bottom panel of Figure \ref{fig:plotall_1} shows the V-band polarization position angle of our precessing disk model compared to the observed polarization position angle after correction for interstellar polarization (following the same method as described in \citealt{mar21}). The minimum polarization position angle of our model is $\sim55.8^{\circ}$ in 1983, and the maximum is $174.2^{\circ}$ in 2021.

Like the H$\alpha$ EW and V-band photometry, the polarization position angle was also observed to rapidly drop in 2007. Prior to this, our precessing disk model closely follows the observed trend, and fits with a reduced $\chi^2 = 1.4$. After the rapid drop our model is unable to reproduce the observations.

\newpage

\subsection{An Ad-hoc Disk Tearing Model} \label{subsec:results_2}

We find that at some times Pleione's H$\alpha$ emitting region must be at a different inclination than the V-band emitting region. For example, the precessing disk model shows that while a single disk can reproduce the rapid drop in H$\alpha$ EW in 2007, the V-band photometry and polarization position angle do not drop at the same time. Based on recent results of \citet{suf22}, we propose an ad-hoc model to reproduce the observed trends. These authors investigate the effects that companion stars have on the dynamics of circumstellar disks. They show that in a Be star-binary system with a 30-day orbital period, where the companion is misaligned by $40^{\circ}$ from the Be star equatorial plane, the disk can separate into two parts with different inclinations while mass is constantly ejected from the stellar surface. In their simulations, the disk periodically tears and merges back into one disk every 30 orbital periods. While their models consider different mass-loss rates, they find that disk tearing only occurs in active phases, i.e., when mass-loss is on. Their results also suggest that disk tearing can occur in a variety of different Be star-binary systems with different masses or orbital periods. Previously, a similar phenomenon was modelled in protoplanetary disks of triple-star systems by \citet{kra20}, who showed the disk around the primary can break into multiple precessing rings aligned with the companion's orbital planes.

We define our ad-hoc model with two distinct stages which occur in each cycle from a Be-shell phase to Be phase and back. Figure \ref{fig:schematic} illustrates these stages, with a) and b) corresponding to the first stage, and c) and d) to the second stage. In essence, the first stage is a single disk model which tilts off-axis, and the second stage is a two-disk model with one disk anchored to the stellar equator while the other disk is free to precess.

In the first stage, the disk is whole and fixed to the stellar equator. Owing to the companion's tidal torque, it becomes tilted away from the midplane by $i\approx30^{\circ}$ (this tilting has also been shown to occur in the simulations of \citealt{suf22} and previously by \citealt{cyr17}). Therefore, in this phase the H$\alpha$ emitting region -- which corresponds to the entire disk, see \citet{car11} -- and the inner disk -- where the V-band flux excess and the optical polarimetry originate -- are approximately aligned.

The second stage begins when the disk is torn into two parts. Now, as shown by the simulations of \citet{suf22} and previously outlined by \citet{oka17}, a gap appears between the small inner disk, which is still anchored to the star, and the outer disk, which starts precessing. Therefore, in this phase the H$\alpha$ emitting region has an orientation that gradually varies in time. Because in this phase the outer disk is no longer fed by the star, it gradually dissipates as the inner disk grows. As a result, the disk slowly transitions to the configuration of the first stage.

The right side of Figure \ref{fig:plotall_1} illustrates our ad-hoc model in detail. In the first stage of our model (indicated by dotted lines in the right panel of Figure \ref{fig:plotall_1}), the innermost disk and H$\alpha$ emitting region have the same inclination as they gradually change from $60^{\circ}$ to $30^{\circ}$. The disk tearing events (1938, 1972, 2007) correspond to the phases immediately before the vertical lines in the figure and mark the beginning of the second stage. Observationally, this stage is related to the Be-shell phase, as the outer disk, now detached from the inner disk, starts precessing around the star. The innermost disk is rebuilt at the original inclination of $60^{\circ}$ over $\sim1.5$ years, with constant mass-injection during this period as suggested from our H$\alpha$ modelling in Subsection \ref{subsec:results_1}. The H$\alpha$ emitting region begins to precess from $30^{\circ}$ to $90^{\circ}$, with a period of $\sim80.5$ years following our best-fit precession model. The precession of the H$\alpha$ region becomes unstable after $\sim15$ years, and the outer disk then gradually dissipates as gas is lost to the interstellar medium or recombines with the innermost disk. As the inner disk grows, it slowly tilts back to $i\approx30^{\circ}$. This marks the phase of the first stage again.

It is important to stress that this ad-hoc model of Pleione's disk addresses the most important observational features seen in the right side of Figure \ref{fig:plotall_1}, namely
\begin{itemize}

\item In the first stage, the gradual change in inclination from $60^{\circ}$ to $30^{\circ}$ of the inner disk explains the rate of change in brightness, the drop in polarization level and the change in the polarization position angle qualitatively.

\item Likewise, the sudden change in the inner disk orientation from $60^{\circ}$ to $30^{\circ}$ once disk tearing has occurred (during the transition from the first stage to the second) also explains the sudden drop in V-band brightness well, and the dramatic change in the polarization position angle (from $\sim120^{\circ}$ in 2007 to $\sim50^{\circ}$ in 2010).

\item The precession model developed in Section \ref{subsec:results_1} remains valid, as in the beginning of the second stage (e.g., the shaded area in the right side of Figure \ref{fig:plotall_1}) the outer disk is allowed to freely precess with a period of $80.5$ yrs.

\end{itemize}

Our ad hoc model, however, is unable to reproduce the strong increase in H$\alpha$ EW seen in the second half of the first stage (e.g., between 1990 and 2007 in the right side of Figure \ref{fig:plotall_1}). At this stage the disk becomes gradually more tilted away from the equator and likely also warped \citep{suf22}. The increase in H$\alpha$ EW might be due to the upper part of the disk, which is closest to the hot stellar poles, becoming hotter and more ionized. This effect, if relevant, is not captured by our simple models that consider axisymmetric and flat disks.

Furthermore, since the density of our disk model was determined solely through H$\alpha$ line profile fitting, the disk density distribution remains constant through the disk tearing period. Perhaps with this scenario, disk gas could be redistributed within the disk which may alter the value of $n$ at particular snapshots in time at certain radial distances from the central star.

In the models of \citet{suf22}, the disk tearing events happen every $\sim30$ orbital cycles for a model with an equal-mass binary in a $30$ day orbit. Since Pleione's less massive companion has a $218$ day orbital period, this means that more than $57$ orbital periods occur before disk tearing every $\sim34$ years.

\section{Discussion and Conclusions} \label{sec:discussion}

We began our work by modelling the large-scale structure of Pleione's disk constrained by H$\alpha$ spectroscopy acquired between 2005 to 2019. During this time, the H$\alpha$ emission shows that Pleione underwent a transition from a Be phase to a Be-shell phase. We find that one single disk model, with fixed density and size, can reproduce the H$\alpha$ observations from the above period by simply varying the disk inclination from $30^{\circ}$ to $80^{\circ}$ at a rate of $\sim 3.7\rm{^{\circ}/year}$. This best-fit disk model is axisymmetric with a base density of $\rho_0 = 3 \times 10^{-11}~\rm{g~cm^{-3}}$, density power-law exponent of $n=2.7$, and a minimum outer disk radius of $R_{out}=15~\rm{R_{eq}}$.

Our H$\alpha$ disk size agrees with \citet{nem10} who found that an upper limit on Pleione's outer disk radius is set by the periastron separation of $53~\rm{R_{\odot}}$ ($12.1~\rm{R_{eq}}$ using the value of $R_{\rm{eq}}$ in Table \ref{tab:28taustellarpara}). However, the outer dimensions of the disk could be restricted by the companion's Roche radius. We note that the lower limit on the outer disk radius, if it is set by the Roche radius, is $\sim7~\rm{R_{eq}}$. We note that in the SPH simulations reported by \citet{cyr17}, they show in tables 3 and 4 that the truncation radius is larger in systems where the disk and companion's orbital plane are not aligned. A full dynamic simulation will be required to determine whether the disk could be further truncated by the companion's Roche lobe.

We fit a precessing disk model to the inclinations determined through H$\alpha$ fitting, and find the best-fit model has a minimum inclination of $25^{\circ}$, a maximum inclination of $145^{\circ}$, and a period of $\sim80.5$ years. We extrapolated this model over 120 years to evaluate the fit to archival observations.

We find that this model reproduces the observed EW from $2001$ to $2021$, including the rapid drop in brightness in $2007$. Our model's peak H$\alpha$ EW is $\sim-2.8~\rm{\AA}$ while the maximum observed value is $\sim-3.8~\rm{\AA}$ in $\sim1960$ and again in $\sim1997$. The observed data from $\sim1970$ to $\sim1985$ could also be matched if the precessing model were shifted by $\sim8$ years, so that $i=30^{\circ}$ in 1971.

Our precessing model is unable to follow the gradual rise in H$\alpha$ EW from $\sim1980$ to $\sim1995$ leading to the peak value, and the decrease from $\sim1997$ to $\sim2001$. The V-band magnitude matches from $\sim1945$ to $\sim1970$ and again from $\sim1990$ to $\sim2007$, but fits poorly outside of these times because of the rapid drops in $1971$ and $2007$. Before $1938$ Pleione was diskless, so the model is not expected to reproduce the observed trend before then. The V-band polarization position angle fits from $\sim1974$ to $\sim2005$, and like the V-band magnitude is unable to capture the rapid drop in $2007$. We note that the model would reproduce the data if it were to reset to $i=60^{\circ}$ in $2007$. We also find the V-band percent polarization fits the observations at all times, but the degree of variation is relatively small.

To align the precessing disk model's polarization position angle to observations, we find that the precession axis of the star must lie at $115^{\circ}$ east of north when projected onto the sky. This is in approximate agreement with the value of $\sim122^{\circ}$ determined by \citet{hir07} with their model based on observations of the polarization position angle.

We further compared our observed polarization position angles to the interferometric disk normal position angle inferred from interferometric observations. Our values were consistent with those of \citet{tou13} who found a value equivalent to $69^{\circ}$ in 2008, and with \citet{coc19} who found a value equivalent to $63^{\circ}$ in 2014 (recall the red dots in the bottom left and right panels of Figure \ref{fig:plotall_1}).

Overall, the precessing model (in which the whole disk precesses with the same rate and orientation) failed to address most of Pleione's observational features.

Inspired by the recent models of \citet{kra20} and \citet{suf22}, we created an ad-hoc disk tearing model which incorporates our precessing disk model and explains the large-scale variations of Pleione's observables, and also seemlessly incorporates the results of \citet{hir07}, \citet{tan07a} and \citet{nem10}. In our ad-hoc model, the disk initially is whole and tilted away from the stellar equator. At some point (likely once the disk becomes sufficiently massive), the disk is torn into two parts: an inner part still anchored to the star, having therefore an inclination angle of $60^{\circ}$, and an outer part which is free to precess around the star. What follows is a slow increase in size and density of the inner disk and a gradual dissipation of the outer disk, which is no longer fed by the central star.

Previously, \citet{hir07} and \citet{tan07a} also found that Pleione's disk inclination changes over time. Using a precessing disk model, \citet{hir07} reproduced the change in polarization position angle across the Be-shell to Be transition in 1989. Using the times between edge-on events, they also found the period of precession to be $80.5$ years, the same value we determined. Building on \citet{hir07}'s model, \citet{tan07a} claimed that the precessing disk had partially re-accreted leading to the rapid drop in brightness in 2007. They also proposed that the sudden appearance of shell lines alongside the drop could be explained by a secondary disk forming in the stellar equatorial plane, which is consistent with our ad-hoc disk tearing model. \citet{nem10} also explained the appearance of shell lines with the formation of a new disk, and we expand on this below.

Our best-fit disk model shows that during the Be to Be-shell transition, that shell lines appear at inclinations near $40^{\circ}$ and greater. We acknowledge that these inclinations are smaller than typically assumed for the appearance of H$\alpha$ shell lines. \citet{han96a} studied the geometry of Be star disks including the range of inclinations typical for shell lines. They find that shell lines most frequently occur for disk inclinations of $75^{\circ} \pm 5^{\circ}$. However, they focused only on changes in geometry and did not consider disk temperature and the associated changes in ionization fraction nor did they consider the density law. Our best-fit model has a disk density slope of $n = 2.7$, meaning that the density falls more gradually with increasing radius, allowing for shell lines to appear at lower inclinations. Also, viscous disks flare with radial distance from the central star (see, e.g., \citealt{car11}) resulting in more material further from the equatorial plane with increasing radial distance. \citet{sil14} also showed that shell lines could form at lower inclinations. For example, for the B8 Ve star, 4 Aql, they modelled shell lines at inclinations of $46^{\circ}$ with $n=2.5$, and at $43^{\circ}$ with $n=3.0$ and $3.5$. Furthermore, in Pleione's tilted disk, the path length through the disk along the line of sight changes over time as the disk inclination varies as well. We also believe that the disk may be warped which could also affect shape of the emission lines although our models only include tilted disks and not warped disks.

We see that when the disk is aligned at the stellar inclination of $61\pm9\rm{^{\circ}}$, the model agrees with each of the observables, suggesting that our disk model is well constrained. \citet{hir07} noted that at the onset of a shell phase in 1973, the disk's inclination was known to be $60^{\circ}$ through polarimetric means. From this, they inferred the star's inclination is also $60^{\circ}$.

We find that our ad-hoc disk tearing model largely reproduces the observed trends of Pleione's circumstellar environment, and agrees with previous descriptions of the disk from the literature. Some details of our model should be examined in further detail, such as the changing thermal structure with a tilting disk, and the dynamical evolution of the disk considering the companion's influence. We also noted a number of other minor discrepancies with work in the literature that we now discuss.

While our model reproduces observations by changing inclination only, we see that small changes to the physical parameters and geometry may help the disk model fit the observables better. In particular, our model suffers from the drawback that the disk is axisymmetric and flat, while the geometry of a precessing disk is more likely curved or warped, with possible density enhancements \citep{mar09}. Because of this, we did not attempt to reproduce the excess H$\alpha$ flux at high velocities ($>180~\rm{km/s}$) observed from 2006~December to 2007~February. These bumps (recall the middle panel in Figure~\ref{fig:halpha_fit_all_phases}), which were also noted by \citet{nem10}, cause an increase in the EW of the line in the transition to the Be-shell phase. At this time, there is a clear transition where the core of the profile decreases while the flux in the wings of the profile increases. This is strongly indicative of a new disk forming at the equator of the star, while at larger radii the disk continues to dissipate. We expect a fully dynamical computation following our ad-hoc disk tearing model would reproduce these features, as after the disk tears the inner disk would continue to contribute to the high-velocity component of the H$\alpha$ line profile.

\citet{sil14} modelled Pleione's disk using the same 2007 Dec 18 H$\alpha$ observation used in this work, and found the disk's density to be $\rho_0 = 6.2 \times 10^{-12}~\rm{g~cm^{-3}}$ and $n=2.5$, at an inclination of $76^{\circ}$. However, we note that our fit is constrained by a series of H$\alpha$ observations and confirmed by V-band photometric and polarimetric observations which show a clear trend, providing a stronger constraint \citep{kle15}.

Pleione is presently in a Be-shell phase which began in 2007. As the inclination of the H$\alpha$ emitting region is once again nearly edge-on, this marks the half-way point to the next rapid drop in brightness. We highly encourage astronomers to continue collecting high time-resolution data on this unique system, to motivate not only the study of Pleione, but other Be stars, such as $\gamma$ Cas and 59 Cyg \citep{hum98}, that have been found to have variable disk inclinations. In addition to studying the large-scale variations caused by the influence of companions on the disk, the impact of smaller scale dynamical changes on the disk, such as those found by \citet{wan17}, should be investigated as well. Future work should also investigate whether SPH simulations using Pleione's orbital parameters are able to explain both the disk tearing period and the outer disk precession period. In our follow-up work, we will determine the effect of disk tilting on the thermal structure of Pleione's disk.

\acknowledgments

The authors would like to thank the referee, Douglas Gies, for comments and suggestions that helped to improve this work. CEJ wishes to acknowledge support through the Natural Sciences and Engineering Research Council of Canada. CT would like to acknowledge, with thanks, FRCE grant support from Central Michigan University. The authors also recognize support from the Lowell Observatory in the form of the telescope time used to obtain the H$\alpha$ line spectra. ACC acknowledges support from CNPq (grant 311446/2019-1) and FAPESP (grants 2018/04055-8 and 2019/13354-1). This work made use of the computing facilities of the Laboratory of Astroinformatics (IAG/USP, NAT/Unicsul), whose purchase was made possible by the Brazilian agency FAPESP (grant 2009/54006-4) and the \mbox{INCT-A}. This work has made use of the BeSS database, operated at LESIA, Observatoire de Meudon, France: http://basebe.obspm.fr. The authors would like to thank John Landstreet from the University of Western Ontario, as well as David Bohlender and Elizabeth Griffin from the Dominion Astrophysical Observatory for helpful discussions in extracting and processing the archival spectrum data. The authors thank Joan Guarro i Fl\'o and Ernst Pollmann for their contribution of a large number of spectra of Pleione to the BeSS database. KCM thanks Amanda Rubio from the Instituto de Astronomia, Geofíscia e Ciências Atmosféricas, at the Universidade de São Paulo, for guidance with \hdust\ and \emcee. KCM is grateful to Emily Pass from the Harvard Smithsonian Center for Astrophysics for identifying the appropriate archival observations located in the Harvard Astronomical Photographic Plate Collection. KCM also thanks Anahi Granada, for helpful discussions on stellar parameterization and evolution.

\bibliography{main}

\begin{thebibliography}{}
\expandafter\ifx\csname natexlab\endcsname\relax\def\natexlab#1{#1}\fi
\providecommand{\url}[1]{\href{#1}{#1}}
\providecommand{\dodoi}[1]{doi:~\href{http://doi.org/#1}{\nolinkurl{#1}}}
\providecommand{\doeprint}[1]{\href{http://ascl.net/#1}{\nolinkurl{http://ascl.net/#1}}}
\providecommand{\doarXiv}[1]{\href{https://arxiv.org/abs/#1}{\nolinkurl{https://arxiv.org/abs/#1}}}

\bibitem[{{Abt} \& {Levato}(1978)}]{hel78}
{Abt}, H.~A., \& {Levato}, H. 1978, \pasp, 90, 201, \dodoi{10.1086/130308}

\bibitem[{{Auer} \& {Mihalas}(1968)}]{aue68}
{Auer}, L.~H., \& {Mihalas}, D. 1968, \apj, 153, 245, \dodoi{10.1086/149654}

\bibitem[{{Binnendijk}(1949)}]{bin49}
{Binnendijk}, L. 1949, \aj, 54, 117, \dodoi{10.1086/106209}

\bibitem[{{Boehme}(1988)}]{boh88}
{Boehme}, D. 1988, Information Bulletin on Variable Stars, 3222, 1

\bibitem[{{Bohme}(1984)}]{boh84}
{Bohme}, D. 1984, Information Bulletin on Variable Stars, 2507, 1

\bibitem[{{Bohme}(1985)}]{boh85}
---. 1985, Information Bulletin on Variable Stars, 2723, 1

\bibitem[{{Bohme}(1986)}]{boh86}
---. 1986, Information Bulletin on Variable Stars, 2893, 1

\bibitem[{{Carciofi}(2011)}]{car11}
{Carciofi}, A.~C. 2011, in Active OB Stars: Structure, Evolution, Mass Loss,
  and Critical Limits, ed. C.~{Neiner}, G.~{Wade}, G.~{Meynet}, \& G.~{Peters},
  Vol. 272, 325--336, \dodoi{10.1017/S1743921311010738}

\bibitem[{{Carciofi} \& {Bjorkman}(2006)}]{car06}
{Carciofi}, A.~C., \& {Bjorkman}, J.~E. 2006, \apj, 639, 1081,
  \dodoi{10.1086/499483}

\bibitem[{{Carciofi} {et~al.}(2007){Carciofi}, {Magalh{\~a}es}, {Leister},
  {Bjorkman}, \& {Levenhagen}}]{car07}
{Carciofi}, A.~C., {Magalh{\~a}es}, A.~M., {Leister}, N.~V., {Bjorkman}, J.~E.,
  \& {Levenhagen}, R.~S. 2007, \apjl, 671, L49, \dodoi{10.1086/524772}

\bibitem[{{Cochetti} {et~al.}(2019){Cochetti}, {Arcos}, {Kanaan}, {Meilland},
  {Cidale}, \& {Cur{\'e}}}]{coc19}
{Cochetti}, Y.~R., {Arcos}, C., {Kanaan}, S., {et~al.} 2019, \aap, 621, A123,
  \dodoi{10.1051/0004-6361/201833551}

\bibitem[{{Cox}(2000)}]{cox00}
{Cox}, A.~N. 2000, {Allen's astrophysical quantities} (New York: AIP Press;
  Springer)

\bibitem[{{Cyr} {et~al.}(2017){Cyr}, {Jones}, {Panoglou}, {Carciofi}, \&
  {Okazaki}}]{cyr17}
{Cyr}, I.~H., {Jones}, C.~E., {Panoglou}, D., {Carciofi}, A.~C., \& {Okazaki},
  A.~T. 2017, \mnras, 471, 596, \dodoi{10.1093/mnras/stx1427}

\bibitem[{{Dapergolas} {et~al.}(1981){Dapergolas}, {di Cola}, {Guarnieri}, \&
  {Madama}}]{dap81}
{Dapergolas}, A., {di Cola}, G., {Guarnieri}, A., \& {Madama}, G. 1981,
  Information Bulletin on Variable Stars, 1920, 1

\bibitem[{{Delplace} \& {Hubert}(1973)}]{del73}
{Delplace}, A.-M., \& {Hubert}, H. 1973, Academie des Sciences Paris Comptes
  Rendus Serie B Sciences Physiques, 277, 575

\bibitem[{{Doazan} \& {Underhill}(1986)}]{doa82}
{Doazan}, V., \& {Underhill}, A. 1986, B Stars With and Without Emission Lines
  (National Aeronautics and Space Administration)

\bibitem[{{Draper} {et~al.}(2014){Draper}, {Wisniewski}, {Bjorkman}, {Meade},
  {Haubois}, {Mota}, {Carciofi}, \& {Bjorkman}}]{dra14}
{Draper}, Z.~H., {Wisniewski}, J.~P., {Bjorkman}, K.~S., {et~al.} 2014, \apj,
  786, 120, \dodoi{10.1088/0004-637X/786/2/120}

\bibitem[{{Ducati}(2002)}]{duc02}
{Ducati}, J.~R. 2002, VizieR Online Data Catalog

\bibitem[{{Dunn} {et~al.}(2006){Dunn}, {Fabian}, \& {Sanders}}]{dun06}
{Dunn}, R.~J.~H., {Fabian}, A.~C., \& {Sanders}, J.~S. 2006, \mnras, 366, 758,
  \dodoi{10.1111/j.1365-2966.2005.09928.x}

\bibitem[{{Egan} {et~al.}(2003){Egan}, {Price}, {Kraemer}, {Mizuno}, {Carey},
  {Wright}, {Engelke}, {Cohen}, \& {Gugliotti}}]{ega03}
{Egan}, M.~P., {Price}, S.~D., {Kraemer}, K.~E., {et~al.} 2003, VizieR Online
  Data Catalog, V/114

\bibitem[{ESA(2000)}]{esa00}
ESA. 2000, INES 3.0 Search Output Description.
\newblock \url{http://sdc.cab.inta-csic.es/ines/OutForm.html}

\bibitem[{{Fitzpatrick}(1999)}]{fit99}
{Fitzpatrick}, E.~L. 1999, \pasp, 111, 63, \dodoi{10.1086/316293}

\bibitem[{{Foreman-Mackey} {et~al.}(2013){Foreman-Mackey}, {Hogg}, {Lang}, \&
  {Goodman}}]{for13}
{Foreman-Mackey}, D., {Hogg}, D.~W., {Lang}, D., \& {Goodman}, J. 2013, \pasp,
  125, 306, \dodoi{10.1086/670067}

\bibitem[{{Freire Ferrero} {et~al.}(2012){Freire Ferrero}, {Morales Dur{\'a}n},
  {Halbwachs}, \& {Cabo Cubeiro}}]{fer12}
{Freire Ferrero}, R., {Morales Dur{\'a}n}, C., {Halbwachs}, J.-L., \& {Cabo
  Cubeiro}, A.~M. 2012, \aj, 143, 28, \dodoi{10.1088/0004-6256/143/2/28}

\bibitem[{{Fr{\'e}mat} {et~al.}(2005){Fr{\'e}mat}, {Zorec}, {Hubert}, \&
  {Floquet}}]{fre05}
{Fr{\'e}mat}, Y., {Zorec}, J., {Hubert}, A.-M., \& {Floquet}, M. 2005, \aap,
  440, 305, \dodoi{10.1051/0004-6361:20042229}

\bibitem[{{Frost}(1906)}]{fro06}
{Frost}, E.~B. 1906, \apj, 23, \dodoi{10.1086/141340}

\bibitem[{{Frost} {et~al.}(1926){Frost}, {Barrett}, \& {Struve}}]{fro26}
{Frost}, E.~B., {Barrett}, S.~B., \& {Struve}, O. 1926, \apj, 64, 1,
  \dodoi{10.1086/142986}

\bibitem[{{Gaia Collaboration} {et~al.}(2020){Gaia Collaboration}, Brown,
  Vallenari, Prusti, de~Bruijne, Babusiaux, \& Biermann}]{gaiaeDR3}
{Gaia Collaboration}, Brown, A. G.~A., Vallenari, A., {et~al.} 2020, Gaia Early
  Data Release 3: Summary of the contents and survey properties.
\newblock \doarXiv{2012.01533}

\bibitem[{{Georgy} {et~al.}(2013){Georgy}, {Ekstr{\"o}m}, {Granada}, {Meynet},
  {Mowlavi}, {Eggenberger}, \& {Maeder}}]{geo13}
{Georgy}, C., {Ekstr{\"o}m}, S., {Granada}, A., {et~al.} 2013, \aap, 553, A24,
  \dodoi{10.1051/0004-6361/201220558}

\bibitem[{{Gies} {et~al.}(1990){Gies}, {McKibben}, {Kelton}, {Opal}, \&
  {Sawyer}}]{gie90}
{Gies}, D.~R., {McKibben}, W.~P., {Kelton}, P.~W., {Opal}, C.~B., \& {Sawyer},
  S. 1990, \aj, 100, 1601, \dodoi{10.1086/115620}

\bibitem[{{Gonz{\'a}lez-Riestra} {et~al.}(2001){Gonz{\'a}lez-Riestra},
  {Solano}, {Talavera}, {Rodr{\'\i}guez}, {Garc{\'\i}a}, {Mart{\'\i}nez},
  {Montesinos}, {Sanz}, {de La Fuente}, {Skillen}, {Ponz}, \&
  {Wamsteker}}]{wam01}
{Gonz{\'a}lez-Riestra}, R., {Solano}, E., {Talavera}, A., {et~al.} 2001, in
  Astronomical Society of the Pacific Conference Series, Vol. 238, Astronomical
  Data Analysis Software and Systems X, ed. J.~{Harnden}, F.~R., F.~A.
  {Primini}, \& H.~E. {Payne}, 156

\bibitem[{{Gulliver}(1977)}]{gul77}
{Gulliver}, A.~F. 1977, \apjs, 35, 441, \dodoi{10.1086/190487}

\bibitem[{{Hall} {et~al.}(1994){Hall}, {Fulton}, {Huenemoerder}, {Welty}, \&
  {Neff}}]{hal94}
{Hall}, J.~C., {Fulton}, E.~E., {Huenemoerder}, D.~P., {Welty}, A.~D., \&
  {Neff}, J.~E. 1994, \pasp, 106, 315, \dodoi{10.1086/133381}

\bibitem[{{Hall} \& {Lockwood}(1995)}]{hal95}
{Hall}, J.~C., \& {Lockwood}, G.~W. 1995, \apj, 438, 404,
  \dodoi{10.1086/175084}

\bibitem[{{Hanuschik}(1996)}]{han96a}
{Hanuschik}, R.~W. 1996, \aap, 308, 170

\bibitem[{{Hirata}(1995)}]{hir95}
{Hirata}, R. 1995, \pasj, 47, 195

\bibitem[{{Hirata}(2007)}]{hir07}
{Hirata}, R. 2007, in Astronomical Society of the Pacific Conference Series,
  Vol. 361, Active OB-Stars: Laboratories for Stellare and Circumstellar
  Physics, ed. A.~T. {Okazaki}, S.~P. {Owocki}, \& S.~{Stefl}, 267

\bibitem[{{Hirata} \& {Kogure}(1976)}]{hir76}
{Hirata}, R., \& {Kogure}, T. 1976, \pasj, 28, 509

\bibitem[{{Hirata} \& {Kogure}(1977)}]{hir77}
---. 1977, \pasj, 29, 477

\bibitem[{{Hopp} \& {Witzigmann}(1980)}]{hop80}
{Hopp}, U., \& {Witzigmann}, S. 1980, Information Bulletin on Variable Stars,
  1782

\bibitem[{{Hopp} {et~al.}(1982){Hopp}, {Witzigmann}, \& {Geyer}}]{hop82}
{Hopp}, U., {Witzigmann}, S., \& {Geyer}, E.~H. 1982, Information Bulletin on
  Variable Stars, 2148

\bibitem[{{Hummel}(1998)}]{hum98}
{Hummel}, W. 1998, \aap, 330, 243

\bibitem[{{Iliev}(2019)}]{ili19}
{Iliev}, L. 2019, IAU Symposium, 346, 149, \dodoi{10.1017/S1743921319002345}

\bibitem[{{Iliev} {et~al.}(2007){Iliev}, {Koubsk{\'y}}, {Kub{\'a}t}, \&
  {Kawka}}]{ili07}
{Iliev}, L., {Koubsk{\'y}}, P., {Kub{\'a}t}, J., \& {Kawka}, A. 2007, in
  Astronomical Society of the Pacific Conference Series, Vol. 361, Active
  OB-Stars: Laboratories for Stellare and Circumstellar Physics, ed. A.~T.
  {Okazaki}, S.~P. {Owocki}, \& S.~{Stefl}, 440

\bibitem[{{Katahira} {et~al.}(1996{\natexlab{a}}){Katahira}, {Hirata}, {Ito},
  {Katoh}, {Ballereau}, \& {Chauville}}]{kat96a}
{Katahira}, J.-I., {Hirata}, R., {Ito}, M., {et~al.} 1996{\natexlab{a}}, \pasj,
  48, 317, \dodoi{10.1093/pasj/48.2.317}

\bibitem[{{Katahira} {et~al.}(1996{\natexlab{b}}){Katahira}, {Hirata}, {Ito},
  {Katoh}, {Ballereau}, \& {Chauville}}]{kat96b}
{Katahira}, J.-I., {Hirata}, R., {Ito}, M., {et~al.} 1996{\natexlab{b}}, in
  Revista Mexicana de Astronomia y Astrofisica Conference Series, Vol.~5,
  Revista Mexicana de Astronomia y Astrofisica Conference Series, ed.
  V.~{Niemela}, N.~{Morrell}, P.~{Pismis}, \& S.~{Torres-Peimbert}, 114

\bibitem[{{Klement} {et~al.}(2015){Klement}, {Carciofi}, {Rivinius},
  {Panoglou}, {Vieira}, {Bjorkman}, {{\v S}tefl}, {Tycner}, {Faes}, {Kor{\v
  c}{\'a}kov{\'a}}, {M{\"u}ller}, {Zavala}, \& {Cur{\'e}}}]{kle15}
{Klement}, R., {Carciofi}, A.~C., {Rivinius}, T., {et~al.} 2015, \aap, 584,
  A85, \dodoi{10.1051/0004-6361/201526535}

\bibitem[{{Kraus} {et~al.}(2020){Kraus}, {Kreplin}, {Young}, {Bate}, {Monnier},
  {Harries}, {Avenhaus}, {Kluska}, {Laws}, {Rich}, {Willson}, {Aarnio},
  {Adams}, {Andrews}, {Anugu}, {Bae}, {ten Brummelaar}, {Calvet}, {Cur{\'e}},
  {Davies}, {Ennis}, {Espaillat}, {Gardner}, {Hartmann}, {Hinkley}, {Labdon},
  {Lanthermann}, {LeBouquin}, {Schaefer}, {Setterholm}, {Wilner}, \&
  {Zhu}}]{kra20}
{Kraus}, S., {Kreplin}, A., {Young}, A.~K., {et~al.} 2020, Science, 369, 1233,
  \dodoi{10.1126/science.aba4633}

\bibitem[{{Lindblad}(1922)}]{lin22}
{Lindblad}, B. 1922, \apj, 55, 85, \dodoi{10.1086/142660}

\bibitem[{{Magalhaes} {et~al.}(1984){Magalhaes}, {Benedetti}, \&
  {Roland}}]{mag84}
{Magalhaes}, A.~M., {Benedetti}, E., \& {Roland}, E.~H. 1984, \pasp, 96, 383,
  \dodoi{10.1086/131351}

\bibitem[{{Magalhaes} {et~al.}(1996){Magalhaes}, {Rodrigues}, {Margoniner},
  {Pereyra}, \& {Heathcote}}]{mag96}
{Magalhaes}, A.~M., {Rodrigues}, C.~V., {Margoniner}, V.~E., {Pereyra}, A., \&
  {Heathcote}, S. 1996, in Astronomical Society of the Pacific Conference
  Series, Vol.~97, Polarimetry of the Interstellar Medium, ed. W.~G. {Roberge}
  \& D.~C.~B. {Whittet}, 118

\bibitem[{{Marr} {et~al.}(2021){Marr}, {Jones}, {Carciofi}, {Rubio}, {Mota},
  {Ghoreyshi}, {Hatfield}, \& {R{\'\i}mulo}}]{mar21}
{Marr}, K.~C., {Jones}, C.~E., {Carciofi}, A.~C., {et~al.} 2021, arXiv
  e-prints, arXiv:2103.06948.
\newblock \doarXiv{2103.06948}

\bibitem[{{Marr} {et~al.}(2018){Marr}, {Jones}, \& {Halonen}}]{mar18}
{Marr}, K.~C., {Jones}, C.~E., \& {Halonen}, R.~J. 2018, \apj, 852, 103,
  \dodoi{10.3847/1538-4357/aaa0d0}

\bibitem[{{Martin} {et~al.}(2009){Martin}, {Pringle}, \& {Tout}}]{mar09}
{Martin}, R.~G., {Pringle}, J.~E., \& {Tout}, C.~A. 2009, \mnras, 400, 383,
  \dodoi{10.1111/j.1365-2966.2009.15465.x}

\bibitem[{{Martin} {et~al.}(2011){Martin}, {Pringle}, {Tout}, \&
  {Lubow}}]{mart11}
{Martin}, R.~G., {Pringle}, J.~E., {Tout}, C.~A., \& {Lubow}, S.~H. 2011,
  \mnras, 416, 2827, \dodoi{10.1111/j.1365-2966.2011.19231.x}

\bibitem[{{McLaughlin}(1938)}]{mcl38}
{McLaughlin}, D.~B. 1938, \apj, 88, 622, \dodoi{10.1086/144014}

\bibitem[{{Merrill} \& {Burwell}(1933)}]{mer33}
{Merrill}, P.~W., \& {Burwell}, C.~G. 1933, \apj, 78, 87,
  \dodoi{10.1086/143490}

\bibitem[{{Mota}(2019)}]{mot19}
{Mota}, B.~C. 2019, PhD thesis, University of São Paulo,
  \dodoi{10.11606/T.14.2020.tde-26052019-143801}

\bibitem[{{Moultaka} {et~al.}(2004){Moultaka}, {Ilovaisky}, {Prugniel}, \&
  {Soubiran}}]{mou04}
{Moultaka}, J., {Ilovaisky}, S.~A., {Prugniel}, P., \& {Soubiran}, C. 2004,
  \pasp, 116, 693, \dodoi{10.1086/422177}

\bibitem[{Murakami {et~al.}(2007)Murakami, Baba, Barthel, Clements, Cohen, Doi,
  Enya, Figueredo, Fujishiro, Fujiwara, Fujiwara, Garcia-Lario, Goto, Hasegawa,
  Hibi, Hirao, Hiromoto, Hong, Imai, Ishigaki, Ishiguro, Ishihara, Ita, Jeong,
  Jeong, Kaneda, Kataza, Kawada, Kawai, Kawamura, Kessler, Kester, Kii, Kim,
  Kim, Kobayashi, Koo, Kwon, Lee, Lorente, Makiuti, Matsuhara, Matsumoto,
  Matsuo, Matsuura, MÜller, Murakami, Nagata, Nakagawa, Naoi, Narita, Noda,
  Oh, Ohnishi, Ohyama, Okada, Okuda, Oliver, Onaka, Ootsubo, Oyabu, Pak, Park,
  Pearson, Rowan-Robinson, Saito, Sakon, Salama, Sato, Savage, Serjeant,
  Shibai, Shirahata, Sohn, Suzuki, Takagi, Takahashi, TanabÉ, Takeuchi,
  Takita, Thomson, Uemizu, Ueno, Usui, Verdugo, Wada, Wang, Watabe, Watarai,
  White, Yamamura, Yamauchi, \& Yasuda}]{mur07}
Murakami, H., Baba, H., Barthel, P., {et~al.} 2007, Publications of the
  Astronomical Society of Japan, 59, S369, \dodoi{10.1093/pasj/59.sp2.S369}

\bibitem[{{Nemravov{\'a}} {et~al.}(2010){Nemravov{\'a}}, {Harmanec},
  {Kub{\'a}t}, {Koubsk{\'y}}, {Iliev}, {Yang}, {Ribeiro}, {{\v S}lechta},
  {Kotkov{\'a}}, {Wolf}, \& {{\v S}koda}}]{nem10}
{Nemravov{\'a}}, J., {Harmanec}, P., {Kub{\'a}t}, J., {et~al.} 2010, \aap, 516,
  A80, \dodoi{10.1051/0004-6361/200913885}

\bibitem[{{Neugebauer} {et~al.}(1984){Neugebauer}, {Habing}, {van Duinen},
  {Aumann}, {Baud}, {Beichman}, {Beintema}, {Boggess}, {Clegg}, {de Jong},
  {Emerson}, {Gautier}, {Gillett}, {Harris}, {Hauser}, {Houck}, {Jennings},
  {Low}, {Marsden}, {Miley}, {Olnon}, {Pottasch}, {Raimond}, {Rowan-Robinson},
  {Soifer}, {Walker}, {Wesselius}, \& {Young}}]{neu84}
{Neugebauer}, G., {Habing}, H.~J., {van Duinen}, R., {et~al.} 1984, \apjl, 278,
  L1, \dodoi{10.1086/184209}

\bibitem[{{Okazaki}(2017)}]{oka17}
{Okazaki}, A.~T. 2017, in The Lives and Death-Throes of Massive Stars, ed.
  J.~J. {Eldridge}, J.~C. {Bray}, L.~A.~S. {McClelland}, \& L.~{Xiao}, Vol.
  329, 432--432, \dodoi{10.1017/S1743921317003143}

\bibitem[{{Parsons}(1918)}]{par18}
{Parsons}, H.~M. 1918, \apj, 47, 38, \dodoi{10.1086/142378}

\bibitem[{{Poeckert} \& {Marlborough}(1979)}]{poe79b}
{Poeckert}, R., \& {Marlborough}, J.~M. 1979, \apj, 233, 259,
  \dodoi{10.1086/157387}

\bibitem[{{Pojmanski}(1997)}]{poj97}
{Pojmanski}, G. 1997, \actaa, 47, 467.
\newblock \doarXiv{astro-ph/9712146}

\bibitem[{{Pollmann}(2011)}]{pol11}
{Pollmann}, E. 2011, BAV Rundbrief - Mitteilungsblatt der Berliner
  Arbeits-gemeinschaft fuer Veraenderliche Sterne, 60, 5

\bibitem[{{Pollmann}(2020)}]{pol20}
---. 2020, BAV Rundbrief - Mitteilungsblatt der Berliner Arbeits-gemeinschaft
  fuer Veraenderliche Sterne, 69, 81

\bibitem[{{Rivinius} {et~al.}(2013){Rivinius}, {Carciofi}, \&
  {Martayan}}]{riv13}
{Rivinius}, T., {Carciofi}, A.~C., \& {Martayan}, C. 2013, \aapr, 21, 69,
  \dodoi{10.1007/s00159-013-0069-0}

\bibitem[{Rohatgi(2020)}]{roha20}
Rohatgi, A. 2020, Webplotdigitizer: Version 4.4.
\newblock \url{https://automeris.io/WebPlotDigitizer}

\bibitem[{{Sharov} \& {Lyutyi}(1988)}]{sha88}
{Sharov}, A.~S., \& {Lyutyi}, V.~M. 1988, \sovast, 32, 303

\bibitem[{{Sharov} \& {Lyutyi}(1992)}]{sha92}
---. 1992, \sovast, 36, 275

\bibitem[{{Sharov} \& {Lyutyi}(1997)}]{sha97}
---. 1997, Astronomy Letters, 23, 93

\bibitem[{{Silaj} {et~al.}(2014){Silaj}, {Jones}, {Sigut}, \& {Tycner}}]{sil14}
{Silaj}, J., {Jones}, C.~E., {Sigut}, T.~A.~A., \& {Tycner}, C. 2014, \apj,
  795, 82, \dodoi{10.1088/0004-637X/795/1/82}

\bibitem[{{Skrutskie} {et~al.}(2006){Skrutskie}, {Cutri}, {Stiening},
  {Weinberg}, {Schneider}, {Carpenter}, {Beichman}, {Capps}, {Chester},
  {Elias}, {Huchra}, {Liebert}, {Lonsdale}, {Monet}, {Price}, {Seitzer},
  {Jarrett}, {Kirkpatrick}, {Gizis}, {Howard}, {Evans}, {Fowler}, {Fullmer},
  {Hurt}, {Light}, {Kopan}, {Marsh}, {McCallon}, {Tam}, {Van Dyk}, \&
  {Wheelock}}]{skr06}
{Skrutskie}, M.~F., {Cutri}, R.~M., {Stiening}, R., {et~al.} 2006, \aj, 131,
  1163, \dodoi{10.1086/498708}

\bibitem[{{Stauffer} {et~al.}(2007){Stauffer}, {Hartmann}, {Fazio}, {Allen},
  {Patten}, {Lowrance}, {Hurt}, {Rebull}, {Cutri}, {Ramirez}, {Young}, {Rieke},
  {Gorlova}, {Muzerolle}, {Slesnick}, \& {Skrutskie}}]{sta07}
{Stauffer}, J.~R., {Hartmann}, L.~W., {Fazio}, G.~G., {et~al.} 2007, \apjs,
  172, 663, \dodoi{10.1086/518961}

\bibitem[{{Suffak} {et~al.}(2022){Suffak}, {Jones}, \& {Carciofi}}]{suf22}
{Suffak}, M., {Jones}, C.~E., \& {Carciofi}, A.~C. 2022, \mnras, 509, 931,
  \dodoi{10.1093/mnras/stab3024}

\bibitem[{{Tanaka} {et~al.}(2007){Tanaka}, {Sadakane}, {Narusawa}, {Naito},
  {Kambe}, {Katahira}, \& {Hirata}}]{tan07a}
{Tanaka}, K., {Sadakane}, K., {Narusawa}, S.-Y., {et~al.} 2007, \pasj, 59, L35,
  \dodoi{10.1093/pasj/59.4.L35}

\bibitem[{{Touhami} {et~al.}(2013){Touhami}, {Gies}, {Schaefer}, {McAlister},
  {Ridgway}, {Richardson}, {Matson}, {Grundstrom}, {ten Brummelaar},
  {Goldfinger}, {Sturmann}, {Sturmann}, {Turner}, \& {Farrington}}]{tou13}
{Touhami}, Y., {Gies}, D.~R., {Schaefer}, G.~H., {et~al.} 2013, \apj, 768, 128,
  \dodoi{10.1088/0004-637X/768/2/128}

\bibitem[{{van Leeuwen}(2007)}]{van07}
{van Leeuwen}, F. 2007, \aap, 474, 653, \dodoi{10.1051/0004-6361:20078357}

\bibitem[{{Wang} {et~al.}(2017){Wang}, {Gies}, \& {Peters}}]{wan17}
{Wang}, L., {Gies}, D.~R., \& {Peters}, G.~J. 2017, \apj, 843, 60,
  \dodoi{10.3847/1538-4357/aa740a}

\bibitem[{{Werner} {et~al.}(2004){Werner}, {Roellig}, {Low}, {Rieke}, {Rieke},
  {Hoffmann}, {Young}, {Houck}, {Brandl}, {Fazio}, {Hora}, {Gehrz}, {Helou},
  {Soifer}, {Stauffer}, {Keene}, {Eisenhardt}, {Gallagher}, {Gautier}, {Irace},
  {Lawrence}, {Simmons}, {Van Cleve}, {Jura}, {Wright}, \&
  {Cruikshank}}]{wer04}
{Werner}, M.~W., {Roellig}, T.~L., {Low}, F.~J., {et~al.} 2004, \apjs, 154, 1,
  \dodoi{10.1086/422992}

\bibitem[{{Wright} {et~al.}(2010){Wright}, {Eisenhardt}, {Mainzer}, {Ressler},
  {Cutri}, {Jarrett}, {Kirkpatrick}, {Padgett}, {McMillan}, {Skrutskie},
  {Stanford}, {Cohen}, {Walker}, {Mather}, {Leisawitz}, {Gautier}, {McLean},
  {Benford}, {Lonsdale}, {Blain}, {Mendez}, {Irace}, {Duval}, {Liu}, {Royer},
  {Heinrichsen}, {Howard}, {Shannon}, {Kendall}, {Walsh}, {Larsen}, {Cardon},
  {Schick}, {Schwalm}, {Abid}, {Fabinsky}, {Naes}, \& {Tsai}}]{wri10}
{Wright}, E.~L., {Eisenhardt}, P. R.~M., {Mainzer}, A.~K., {et~al.} 2010, \aj,
  140, 1868, \dodoi{10.1088/0004-6256/140/6/1868}

\end{thebibliography}

\end{document}